\documentclass[twocolumn]{aastex631}
\usepackage{comment}

\shorttitle{Flares Big and Small}
\shortauthors{Zeldes et al.}
\graphicspath{{./}{figures/}}

\begin{document}

\title{Flares Big and Small: a K2 and TESS View of ASAS-SN Superflares}

\author[0000-0003-3909-8562]{Jesse Zeldes}
\affiliation{Institute for Astronomy, University of Hawai`i, 2680 Woodlawn Dr. Honolulu, HI 96822, USA }
\affiliation{Haverford College, 370 Lancaster Ave., Haverford, PA, 19041, USA}

\author[0000-0001-9668-2920]{Jason T. Hinkle}
\affiliation{Institute for Astronomy, University of Hawai`i, 2680 Woodlawn Dr. Honolulu, HI 96822, USA }

\author[0000-0003-4631-1149]{Benjamin J. Shappee}
\affiliation{Institute for Astronomy, University of Hawai`i, 2680 Woodlawn Dr. Honolulu, HI 96822, USA }

\author[0000-0003-1719-5046]{Ellis A. Avallone}
\affiliation{Institute for Astronomy, University of Hawai`i, 2680 Woodlawn Dr. Honolulu, HI 96822, USA }

\author{Sarah J. Schmidt}
\affiliation{Leibniz-Institute for Astrophysics Potsdam (AIP), An der Sternwarte 16, D-14482, Potsdam, Germany}

\author{Jennifer L. van Saders}
\affiliation{Institute for Astronomy, University of Hawai`i, 2680 Woodlawn Dr. Honolulu, HI 96822, USA }

\author{Zachary Way}
\affiliation{Department of Astronomy, The Ohio State University, 140 West 18th Avenue, Columbus, OH 43210, USA}

\author[0000-0001-6017-2961]{Christopher S. Kochanek}
\affiliation{Department of Astronomy, The Ohio State University, 140 West 18th Avenue, Columbus, OH 43210, USA}
\affiliation{Center for Cosmology and Astroparticle Physics, The Ohio State University, 191 W.~Woodruff Avenue, Columbus, OH 43210, USA}

\author[0000-0001-9206-3460]{Thomas W.-S. Holoien}
\affiliation{The Observatories of the Carnegie Institution for Science, 813 Santa Barbara St., Pasadena, CA 91101, USA}



\begin{abstract}

 We investigate the flare-frequency distributions of 5 M-dwarfs that experienced superflares with energies in excess of \(10^{33}\) erg detected by ASAS-SN. We use K2 and TESS short-cadence observations along with archival ASAS-SN data to categorise the flaring behaviour of these stars across a range of flare energies. We were able to extract a rotation period for 4 of the stars. They were all fast rotators (\(P_{\mathrm{rot}} \leq 6 \textrm{d}\)), implying relative youth. We find that the flare-frequency distributions for each of the stars are well fit by a power-law, with slopes between \(\alpha = 1.22\) and \(\alpha= 1.82\). These slopes are significantly flatter than those of fast-rotating M-dwarfs not selected for their superflaring activity, corresponding to an increased number of high energy flares. Despite our specific selection of superflaring stars with shallow flare-rate distributions and more power in higher-energy flares, we find that the implied UV flux is insufficient to deplete the ozone of earth-sized planets in the habitable zone around these stars.  Furthermore, we find that the flares detected on the stars in our sample are insufficient to produce the UV flux needed to fuel abiogenetic processes.  These results imply that given available models, even M-dwarfs selected for extreme flaring properties may have insufficient UV emission from flares to impact exolife on earth-sized planets in the habitable zones around M-dwarfs.

\end{abstract}



\section{Introduction}
M-dwarfs (stars of effective temperature \(2300\)K to \(3800\)K) are some of the most magnetically active stars \citep[e.g.,][]{1985BAAS...17..751S, Browning_2008}. This extreme magnetism manifests both through quiescent, steady-state,  chromospheric emission \citep[e.g.][]{2012ApJ...748...58D} and through more extreme events such as flares: sudden, multi-wavelength jumps in brightness caused by magnetic field lines reconnecting \citep{Priest2002}. These flares can provide a rare window into the underlying magnetic dynamics of these stars, which are otherwise hard to observe. 

Though stellar magnetic fields driven by a dynamo process typically evolve slowly, during a flare there is a rapid change in the local magnetic field configuration  \citep[e.g.,][]{Priest2002,680140810}. This change accelerates charged particles in the star's upper atmosphere into the surface, temporarily heating a section of the stellar surface to a temperature of $\sim 8,000 - 14000$ K \citep[e.g.,][]{680140843}. The radiation from this hot spot, along with radiation from atmospheric heating and acceleration is then observed as a flare. This process of reconnection can also create beams of charged particles that are emitted outward from the star \citep[e.g.,][]{Priest2002}. 
Because of their intense global magnetism, flares on M-dwarfs are both more frequent and more energetic than those on the Sun, both in equivalent duration (ED) -- a measure of how long in quiescence it would take a star to emit the energy of a flare -- and in total energy emitted \citep[e.g][]{Davenport_2014,2014ApJ...797..121H}. 

Understanding the flaring of these stars is also crucial to studies of exoplanet habitability \citep[e.g.,][]{Grootel_2018}. It is estimated that up to 70\% of habitable zone planets orbit M-dwarfs \citep[e.g.,][]{2015ApJ...807...45D, ilan_2018}.  However, the habitable zones of M-dwarfs are very close to the star, much less than 1 AU, due to their low luminosities \citep[e.g.,][]{680140849,2013ApJ...765..131K}. This means that UV emission and ionised particles emitted during flares could deplete the atmospheres of planets in the M-dwarf's habitable zone, rendering them inhospitable \citep[e.g.,][]{680140839,680140862}. This is particularly true for the highest energy flares, which simulations suggest could be more damaging than an equivalent magnetic energy released in multiple smaller flares, due to higher UV and proton fluxes having a larger impact on exoplanet atmospheric structure. \citep[e.g.,][]{2016PhR...663....1S, Tilley2019}.

Observed M-dwarf flares typically consist of three phases: a rapid, minute-scale, quartic polynomial rise toward the maximum flux, an initial minutes-long polynomial decay phase, and an hour-scale exponential decay \citep{Davenport_2014}. The flare duration scales roughly linearly with the peak flux during the flare \citep{680140826}, though the relationship is different for simple flares with a single peak and complex multi-peaked flares \citep{680140866}.

Flaring M-dwarfs typically flare stochastically and frequently, often multiple times per day, with more flares at lower energies \citep[e.g.,][]{ 2012ApJ...748...58D, 2014ApJ...797..121H,Paudel_2018}. The exact relationship between flare energy and frequency has been a subject of intense study. This relationship is typically quantified by the flare-frequency distribution (FFD), which measures the rate of flares as a function of energy. FFDs typically follow a power-law distribution \cite[e.g,][]{GERSHBERG1973,680140844,gunther} of the form \(f=\beta (E/E_\mathrm{base})^{-\alpha}\), where \(f\) is the frequency in flares per day, \(E\) is the flare energy, and \(E_\mathrm{base}\) is the baseline energy around which the power-law is fit. The spectral index, \(\alpha\), is typically within the range \(1.4\) to \(2.4\), with significant variance across studies \citep[e.g.,][]{2011PhDT.......144H, ilan_2018, 2019ApJ...881....9H}.

Some M-type dwarfs have also been observed to have superflares with energies in excess of \(10^{33}\) erg, which often manifest as a brightening on the scale of \(8-11\) mag depending on contrast \citep[e.g.][]{Schmidt_2019,680140850}. \citet{Paudel_super} suggest that these types of flares likely appear most often on young, fast-rotating stars due to the rotation driving a more powerful magnetic dynamo. However, these results are in tension with other studies, such as \citet{Mondrik_2018}, who find lower flaring rates in fast rotating (period  \(<10\)d) stars than at intermediate (\(10-70\)d) rotation periods.

The stochastic nature of flares presents a challenge for traditional short-duration ground-based observing campaigns \citep{680140844}.  Even resource-intensive, multi-night monitoring campaigns may only yield a few dozen flares per star, and often none of the highest energy flares \citep{680140898}. Space-based and ground-based survey missions, often designed with other primary science goals, have proven to be essential for understanding flares.

Space-based missions like \textit{Kepler} and TESS have much longer continuous baseline observations for each star, opening the door for in-depth studies of flaring behaviour. \textit{Kepler} was an exoplanet-finding mission which monitored 200,000 stars looking for planetary transits \citep{Borucki}. \textit{Kepler}, and its follow-up mission K2, \citep{680140837} have proved crucial tools for studying flares \citep[e.g.][]{2014ApJ...797..121H,Davenport_2014, Gizis_2017, Mondrik_2018, 2019ApJ...871..241D}. Constant monitoring of targets has enabled reliable measurements of flares across many orders of magnitude in energy. \textit{Kepler}'s one-minute short-cadence mode has also made it possible to explore the energy distributions of flares, allowing for the first complete classifications of the flaring behaviour of stars other than the Sun (e.g., \citealp{Paudel_super,Notsu_2013,ilan_2018}).

The successor mission to K2, the Transiting Exoplanet Survey Satellite (TESS, \citealp{2014SPIE.9143E..20R}) has also proved a valuable tool for observing M-dwarf flares \citep[e.g.,][]{gunther, 2020arXiv200510281D, Feinstein_2020}. Unlike \textit{Kepler}, which focused on finding planets around Sun-like stars, TESS's primary science mission seeks to observe planetary transits around bright, nearby low-mass dwarfs. To date, TESS has observed over 200,000 stars in its 2-minute short cadence mode. 

While the long baseline and fine precision of space telescopes are unparallelled for studying individual stars, automated ground-based surveys can search for flares across many more targets simultaneously \citep[e.g.,][]{680140841,680140831,680140823}, providing an essential complement to the space-based missions. One such survey, the All-Sky Automated Survey for Supernovae (ASAS-SN, \citealp{680140864, 680140840}) monitors the entire visible sky to a depth of \(g \approx 18\) mag with a cadence of \(\sim\) 20 hours, with a goal of detecting nearby supernovae. The regular cadence and large search area have enabled ASAS-SN to discover many large flares from L- and M-type dwarfs \citep{680140870,680140886, 680140867, Schmidt_2019}. \citet{Schmidt_2019} and \citet{Rodriguez_Mart_nez_2020} carried out a detailed analysis of the flares in the ASAS-SN \(V\)-band dataset. 

ASAS-SN data improved our understanding of the frequency of superflares and the stars on which they originate, but it is only able to detect the flares with the highest energies for a large sample of M-dwarfs. This makes it difficult to categorise the properties of more typical flares on the stars where superflaring was observed. In this paper, we combine archival ASAS-SN observations with K2 and TESS short-cadence observations to examine 5 M-dwarfs that had superflares observed by ASAS-SN. The details of the ASAS-SN, K2, and TESS data are described in Section \ref{methods}, along with our flare-finding techniques. In Section \ref{Gyrochronology}, we attempt to determine the rotation rate of each of the stars. In Section \ref{sec:FlareDist}, we look at the flare-frequency distribution of each of the stars. Section \ref{discussion} provides a discussion of the results.

\section{Flare Detection}
\label{methods}
In this section, we outline our procedure for identifying candidate superflaring stars, finding flares, and recovering frequencies and bolometric energies for each event. Section \ref{ASSSNpipe} describes the data from the ASAS-SN survey and our target selection criteria. Section \ref{ASAS-recov} outlines our flare-finding techniques for the ASAS-SN data. In Section \ref{K2pipe} we outline our processing and detrending of our space telescope data. In Section \ref{FindFlare}, we describe our flare-finding algorithm based on \texttt{AltaiPony} \citep{2020arXiv201005576I}, a python-based package for automated flare finding. In Section \ref{Injection}, we explain our process for injecting and recovering model flares, which we use to correct the flare energy distributions and estimate our detection efficiency. Finally, in Section \ref{FlareEnergy} we outline our procedure for estimating bolometric energies on identified K2 and TESS flares.

\subsection{ASAS-SN Pipeline and Sample Selection}
\label{ASSSNpipe}

The superflares that motivated this study, with the exception of the flare on ASASSN-20gu, were observed when the ASAS-SN project still consisted of two mounts, which monitored the sky to a depth of \(V\sim 17\) mag every few days. For each of the stars in our sample, we searched the ASAS-SN $g$- and \(V\)-band data for additional flare candidates.

For this paper, we consider any star which was observed to have a flare in ASAS-SN with a recovered energy greater than \(10^{33}\) erg (as estimated by \citealp{Schmidt_2019}), that was also observed in short-cadence mode by either TESS or K2 for at least one sector or quarter, respectively. This gave us a total of 4 stars: the M4 dwarfs ASASSN-13cm, -14jy, and -16dj, observed by TESS, and the M6 dwarf ASASSN-14mz, observed by K2. We also examine the M3 dwarf ASASSN-20gu, which recently superflared in ASAS-SN and for which TESS short cadence data was available. Table \ref{tab:targets} gives the observational details for each of the targets, along with the stellar properties. 

The ASAS-SN data were reduced using a pipeline based on the ISIS image subtraction package \citep{alard98, alard00}. Each epoch typically consists of three dithered 90-second exposures. Each exposure is individually subtracted from a high-quality reference image. We use the IRAF package \texttt{apphot} to perform aperture photometry using a 2-pixel (approximately $16.\!\!''0$) radius aperture on subtracted images, generating a differential light curve of the star. We then add back in the flux from the reference image using the same aperture. This enables us to produce a higher quality light curve than standard aperture photometry methods.  The photometry was calibrated using the AAVSO Photometric All-Sky Survey \citep{henden15}. All low-quality ASAS-SN images were inspected by-eye, and images containing clouds or other systematic problems were removed.  

\begin{deluxetable*}{cccccccc}[htbp!]
\tablecaption{Sample of M-dwarfs}
\tablehead{
\colhead{Target} &
\colhead{Spectral Type} &
\colhead{Energy} &
\colhead{Observation} &
\colhead{Quarter/Sector} &
\colhead{Right Ascension} &
\colhead{Declination} &
\colhead{Distance}\\
&
&
\colhead{(log[E/erg]))}
&
&
& 
&
&
\colhead{(pc)}}
\startdata
ASASSN-13cm & M4 & 33.5 & TESS & 3 & 01 46 51.4 & $-$16 52 19.7 & 59.1 $\pm$ 0.2 \\ 
ASASSN-14jy & M4 & 34.2  & TESS & 1-4, 6-13 & 07 06 58.9 & $-$62 21 10.9 & 46.3 $\pm$ 0.1 \\
ASASSN-14mz & M6 & 33.8 & K2 & 16 & 08 51 13.9 & +19 12 21.5 & 74.4 $\pm$ 13.5 \\
ASASSN-16dj & M4 & 33.9 & TESS & 14, 20, 21 & 10 07 17.7 & +69 20 46.2 & 52.5 $\pm$ 0.1 \\
ASASSN-20gu & M3 & 33.4& TESS & 14 & 19 35 29.1 & +37 46 06.2 & 14.4 $\pm$ 0.01 \\
\enddata 
\tablecomments{The observation properties of each M-dwarf. The energy of the triggering superflare in ASAS-SN is estimated using the template fit described in \ref{ASASFlare}.}
\label{tab:targets}
\end{deluxetable*}

\subsection{ASAS-SN Flare Recovery}
\label{ASAS-recov}
The typical ASAS-SN observation strategy of 2 to 3 dithers does not provide enough data points for a detailed analysis of the flare lightcurve. However, we are still able to make an estimate of flare energy. In addition to the triggering flare and subsequent flares analysed in \citet{Schmidt_2019}, we search for additional flares in each ASAS-SN lightcurve by looking for observations in which there are two or more points above the \(3\sigma\) detection threshold in a given night. We then verify these events by eye.

\begin{figure}
    \centering
    \includegraphics[width=\linewidth]{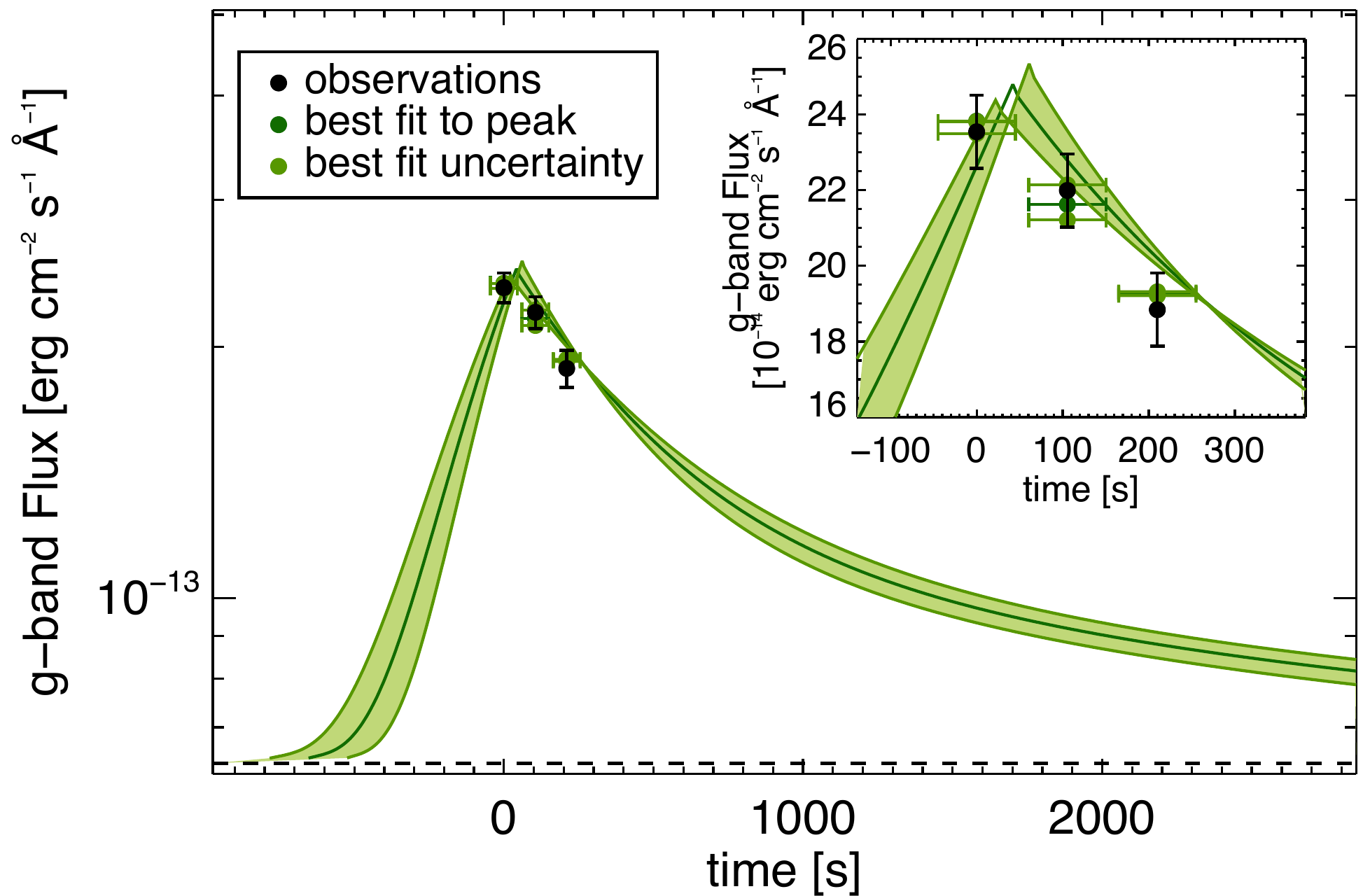}
    \caption{An example of the flare fitting technique for the triggering superflare on ASASSN-20gu. The 3 ASAS-SN data points, each gathered near the flare peak, are shown in black. The inset shows the same data near the peak. The black dashed line shows the quiescent flux value. The dark green line shows the best fit to the flare template, while the light green area shows the range of flare models consistent with the data.}
    \label{fig:asas-flare}
\end{figure}

Once we identity the flares, we calculate an ED estimate for each by fitting a flare using the flare template from \citet{Davenport_2014}. We restrict the modeled flare peak to be within \(2000\) s of the ASAS-SN observations, and the characteristic time scale \(t_{1/2}\) to be within the range \(10\textrm{s} \leq t_{1/2} \leq 2000 \textrm{s}\). We make a grid of possible distributions of \(t_{1/2}\) and peak location, then integrate each over the ASAS-SN 90 second exposure time \citep{Schmidt_2019}. We consider each combination of \(t_{1/2}\) and peak location that agrees with each of our observed ASAS-SN fluxes to within \(1\sigma\) to be a possible flare parameterisation. We then integrate each flare lightcurve to compute the the ED of the event in the ASAS-SN band, given in Table \ref{tab:flares}. We report the median of these EDs to be our most likely value, and the 1st and 3rd quartiles to be our asymmetric error bars. An example of the fitting procedure for the triggering superflare on ASASSN-20gu is shown in Figure \ref{fig:asas-flare}. In total, we identify 27 additional flares for the five stars observed by ASAS-SN. 

\subsection{K2 and TESS Pipelines}
\label{K2pipe}

 Once we have selected our flaring targets from ASAS-SN, we use short cadence data from the space telescopes to examine their flaring behaviour at lower energies. We obtain K2 data for ASASSN-14mz through the K2 guest observer programme (Investigation ID: GO16104\_SC, PI: Shappee).  We detrend these lightcurves using \texttt{K2SC} \citep{Aigrain}, which allows the removal of K2 systematics to almost the level of precision of the original \textit{Kepler} mission. \texttt{K2SC} uses Gaussian processes to model the telescope pointing jitter introduced by K2's periodic altitude corrections using the solar wind. It uses correlations across pixels in the K2 field to correct for time and position-dependent systematics while keeping the real variability. We split the lightcurve into 4 sections using the recommended \texttt{K2SC} splits, and detrend each section separately. 

Our four TESS targets were each selected for short cadence observations based on the TESS mission science criteria outlined in \citet{2014SPIE.9143E..20R}. For our TESS analysis, we use the \texttt{PDCSAP FLUX} lightcurves, which we obtain directly from the TESS mission. We require \texttt{flag quality} $=0$, to ensure the highest-fidelity light curves. The TESS data are detrended using a 3rd order Savitzky–Golay filter, combined with the sigma clipping approach described in \citet{2016ApJ...829...23D}. This enables the smoothing of the data to remove small scale systematics with little loss of true variability.

\subsection{K2 and TESS Flare Identification}
\label{AltaiPony}

Once we have cleaned our data, the next step is identifying possible flare candidates. The detrending and flare-finding algorithms for our TESS and K2 observations are done using \texttt{AltaiPony}\footnote{https://github.com/ekaterinailin/AltaiPony/}, a python-based package for the analysis of flares in \textit{Kepler}, K2, and TESS data. \texttt{AltaiPony} \citep{2020arXiv201005576I} was developed as a successor to the flare-finding software \texttt{appaloosa} \citep{2016ApJ...829...23D}.
Both the K2 and TESS detrending procedures are implemented using the \texttt{FlareLightCurve Detrend} method.

\label{FindFlare}
We found a preliminary sample of flare candidates using the \texttt{AltaiPony find\_flares} method, which implements the \texttt{FINDflare} algorithm defined in \citet{2015ApJ...814...35C}. For a flux \(f_i\) at the \(i\)th epoch of a light curve segment \(L\) to be categorised as a candidate flare, we require that it pass 3 criteria. First, it must satisfy 
\begin{equation}
\frac{\left|f_{\mathrm{i}}-\bar{f}_{\mathrm{L}}\right|}{\sigma_{\mathrm{L}}} \geq N_{1} 
\end{equation}
where \(\bar f_L\) is the mean of the local light curve segment, and \(\sigma_L\) is the standard deviation of the fluxes. To account for photometric uncertainties we also require
\begin{equation}
\frac{\left|f_{\mathrm{i}}-\bar{f}_{\mathrm{L}}+\omega_{\mathrm{i}}\right|}{\sigma_{\mathrm{L}}} \geq N_{2}
\end{equation}
where \(\omega_i\) is the photometric uncertainty at epoch \(i\). Finally, we require that 
\begin{equation}
N_{dat} \geq N_{3}.
\end{equation}
  where \(N_{dat}\) is the number of data points in the candidate flare.  We choose \(N_3 = 4\) for our 1-minute cadence K2 data and \(N_3=2\) for our 2-minute cadence TESS data, yielding a minimum detectable flare length of \(4\) minutes for each of our targets. Following \citet{ilan_2018}, we choose \(N_1=3,N_2=4\). We tested a number of other parameter configurations, and find that these choices seem best able to effectively recover the events that appear to be flares by eye while minimising false positives due to instrumental effects. For each detected flare, we compute the ED of the event in the \textit{Kepler} or TESS band in the same manner as for the ASAS-SN flares. 

\begin{figure}
    \centering
    \includegraphics[width=\linewidth]{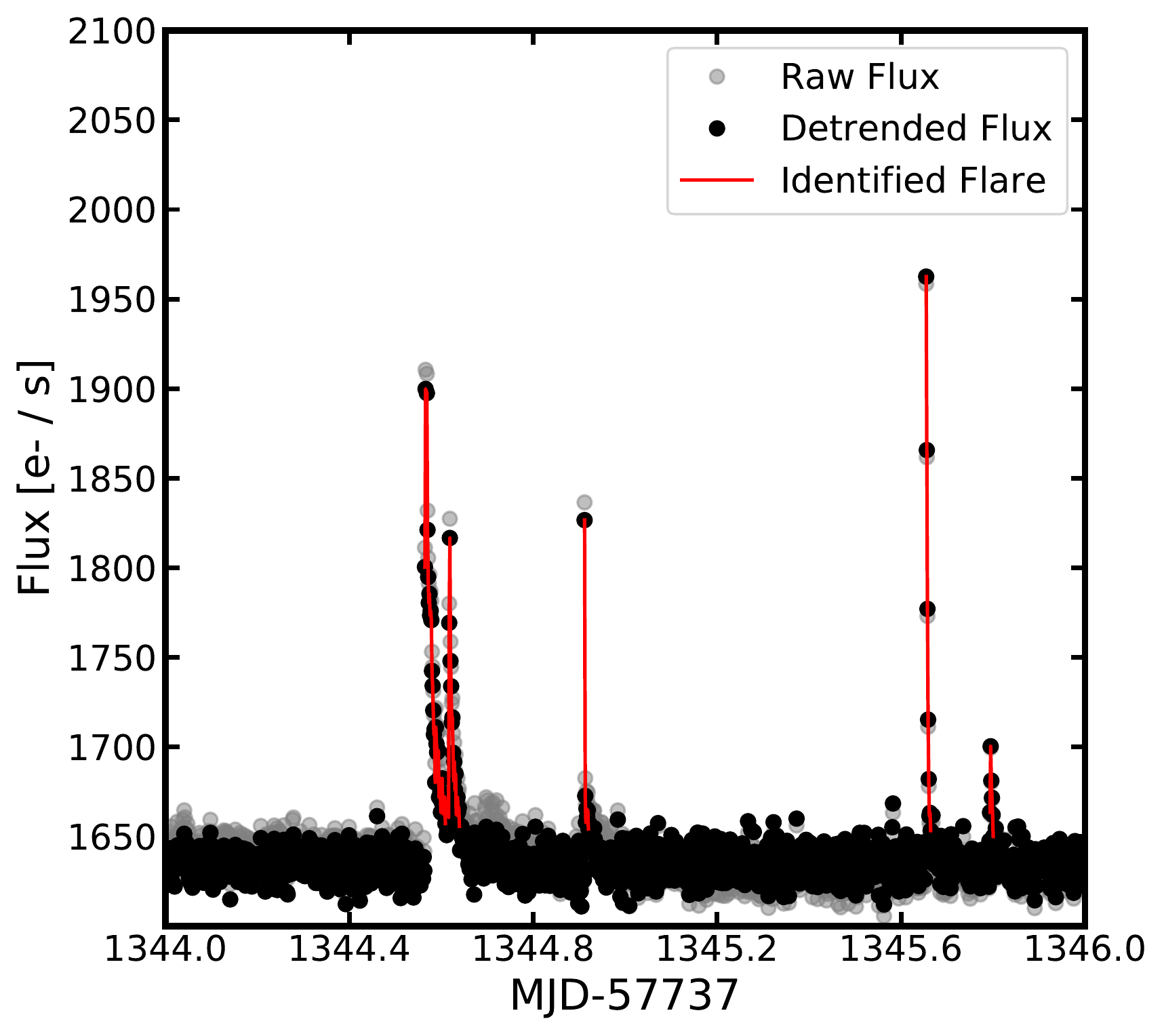}
    \caption{An example of a 2-day section of a TESS lightcurve of ASASSN-14jy. The raw flux is in grey, while the detrended flux is in black. The flares detected by our FINDflare algorithm are highlighted in red.  The flares occur stochastically across the lightcurve and span a range of energies.}
    \label{fig:LC}
    \label{fig:my_label}
\end{figure}

\begin{figure}
    \centering
    \includegraphics[width=\linewidth]{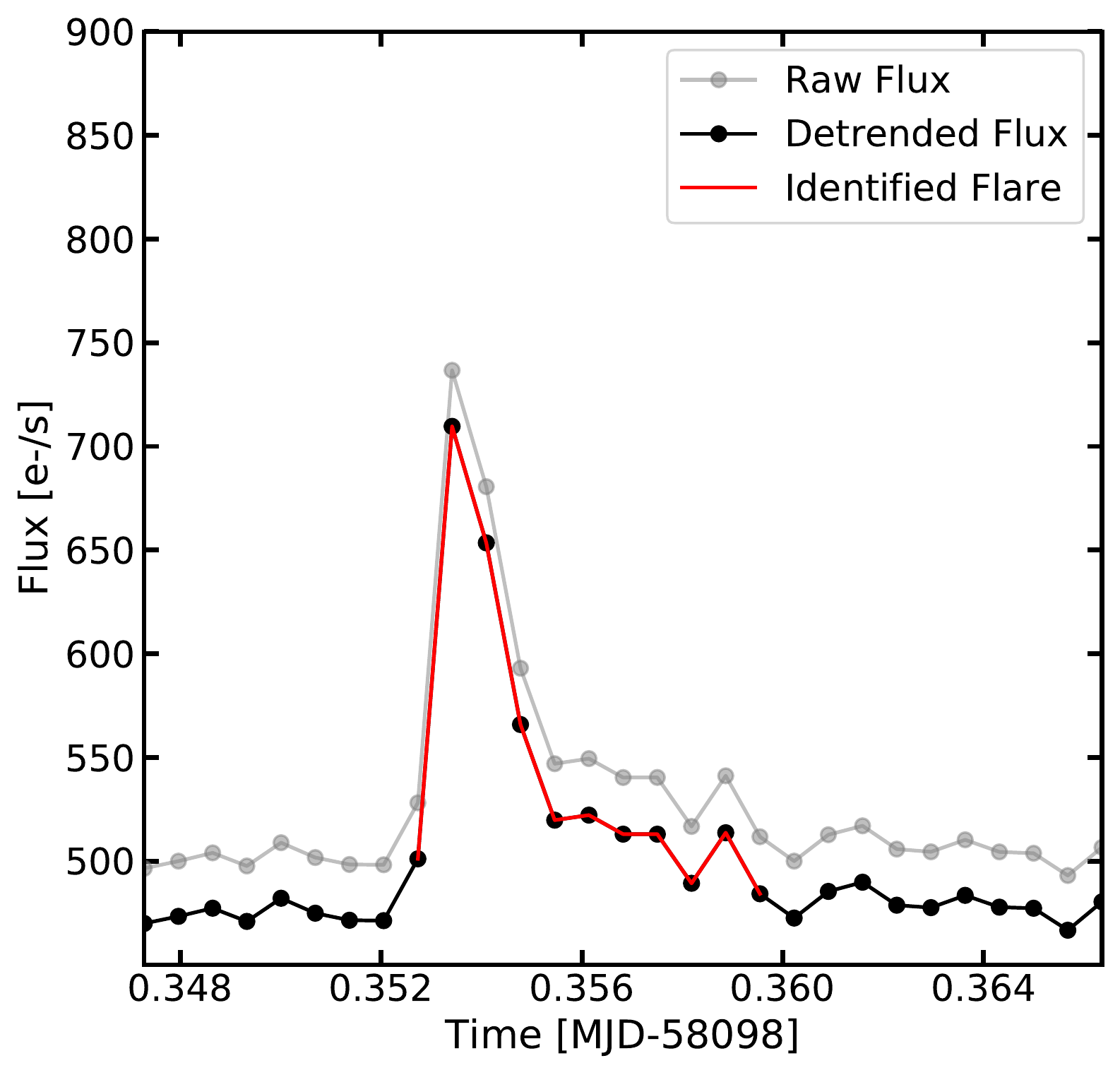}
    \caption{A flare for the K2 target ASASSN-14mz. The raw flux is shown in grey, while the detrended flux is in black. The red outline shows the points that the \texttt{FINDFlare} algorithm classified as part of the event.}
    \label{fig:flare}
\end{figure}

Figure \ref{fig:LC} shows a portion of the TESS data for ASASSN-14jy. The \texttt{FINDflare} algorithm identified 5 flares over this approximately 2-day section of data. The flares, highlighted in red, span a range of energies. Other events, where only a single elevated point is visible, are not categorised as flares. Though it is possible that these points are flares, the lack of multiple detections makes it impossible to identify the classic flare shape and distinguish these points from other kinds of background noise. Figure \ref{fig:flare} gives an example flare identified by the \texttt{FINDflare} algorithm in the K2 light curve of ASASSN-14mz. The characteristic shape of the flare, consisting of an impulsive rise followed by an exponential decay, is clearly evident. This flare has an approximate duration of 8 minutes, and an uncorrected ED of 92.5s. Table \ref{tab:flares} lists all of the flares detected for any of our targets. The time of detection, and the estimated energies are given. In total, we detect 206 flares across the 5 stars in our K2 and TESS data.

\begin{deluxetable*}{cccccccc}[htbp!]
\tablecaption{Observed Flares}
\tablehead{
\colhead{Start Time} &
\colhead{End Time} &
\colhead{Target} &
\colhead{Telescope} &
\colhead{Recovery Probability} &
\colhead{Integrated ED} &
\colhead{Corrected ED} &
\colhead{Corrected Energy}\\
 \colhead{(MJD-57737)} &
 \colhead{(MJD-57737)} &
&
&
\colhead{(s)}& 
\colhead{(s)}&
&
\colhead{(log[E / erg]))}}
\startdata
    1652.05 &   1652.05 &  ASASSN-14jy &      TESS &        $0.42 \pm 0.05$ &     $5.93 \pm 1.19$ &    $6.77 \pm 1.03$ &   $30.48^{+0.12}_{-0.16}$ \\
    1548.05 &   1548.05 &  ASASSN-14jy &      TESS &        $0.24 \pm 0.04$ &     $6.81 \pm 1.09$ &     $6.85 \pm 1.40$ &    $30.5^{+0.15}_{-0.23}$ \\
    1522.41 &   1522.42 &  ASASSN-14jy &      TESS &        $0.73 \pm 0.04$ &     $6.21 \pm 1.08$ &    $7.13 \pm 0.83$ &   $30.51^{+0.09}_{-0.12}$ \\
    1669.30 &   1669.30 &  ASASSN-14jy &      TESS &        $0.92 \pm 0.03$ &     $7.22 \pm 1.31$ &    $8.07 \pm 0.83$ &    $30.56^{+0.08}_{-0.1}$ \\
    1642.12 &   1642.12 &  ASASSN-14jy &      TESS &        $0.73 \pm 0.04$ &     $7.12 \pm 1.31$ &    $8.17 \pm 0.95$ &   $30.56^{+0.09}_{-0.12}$\\
    & & & & \vdots & & &\\
\enddata 
\tablecomments{Example flares for the target ASASSN-14jy. For each flare, we give the starting and ending time, integrated ED, and our final recovered EDs and energies. A full machine readable version of the table containing all of the flares for each of our targets is available.}
\label{tab:flares}
\end{deluxetable*}

\subsection{Injection and Recovery}
\label{Injection}
For each of our targets in K2 and TESS, we inject model flares into the light curve and try to recover them. This injection and recovery analysis serves two purposes: (1) it quantifies how our flare finding efficiency drops off at low energies, and (2) we can use the differences between the injected and recovered flare samples to correct for flux removed by the detrending process and the sampling cadence. For each TESS/K2 section for each star, we inject 50,000 flares over $\sim 100$ trials, ensuring that injected flares are spaced apart by at least 4x the length of the previous flare and that they do not overlap with flares already in the data. The injected flare EDs are sampled from a power-law given by a preliminary fit to the ED-frequency relation.
The template for each of the injected flares in generated using \texttt{AltaiPony}'s \texttt{sample\_flare\_recovery} method, which generates the flares based on the template in \citet{Davenport_2014}. For TESS targets, we run our detrending procedure after injection before running our standard \texttt{FINDflare} recovery procedure. This was not possible for our K2 target due to computational limitations. For stars with light curves from multiple sectors, we analyze each sector separately to account for differing systematics.  A flare was considered recovered if the true flare peak time was contained within the start and end times estimated for the detection.

A weakness of this method is that it assumes that all flares follow the same basic template, and this template is built from data from a single star, the M4 dwarf GJ 1243. If flares on other stars are morphologically different this may be a poor assumption. Additionally, when developing the template, \citet{Davenport_2014} found that \(\approx 15\%\) of flares were complex events, which were not well fit by the simple template, and that these complex events are more common at higher energies. Since flare recovery rates approach 100\% at higher energies, regardless of flare shape, these high energy differences are less important. Quantifying the effect of the flare template shape on recovery rate is beyond the scope of this paper, although we note that, because of the low fraction of complex flares, it is unlikely to cause an error much greater than 15\%.

\begin{figure}
    \centering
    \includegraphics[width=\linewidth]{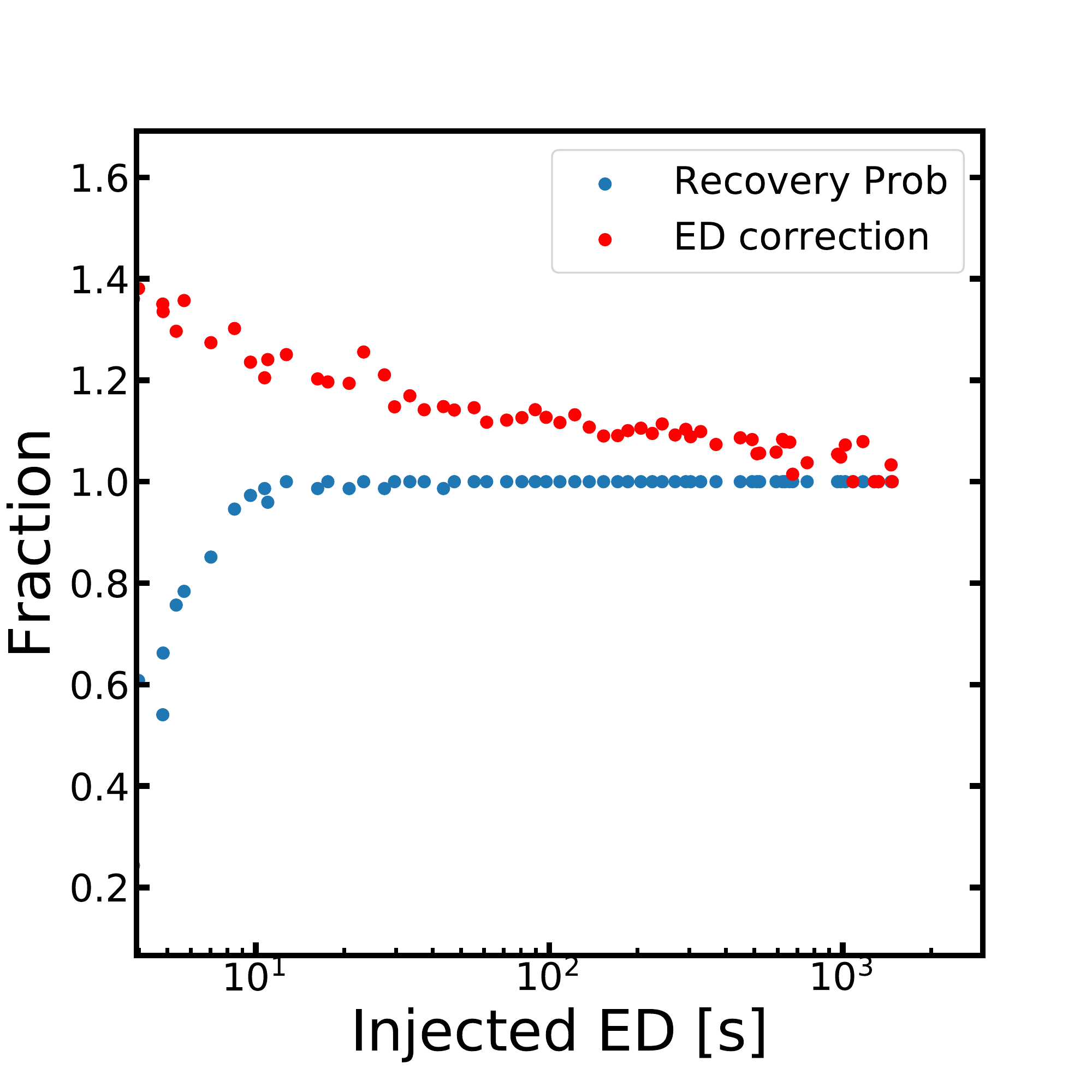}
    \caption{Recovery probability (blue) and ED correction (red) for the  recovery process as a function of the flares' injected ED for the target ASASSN-14jy. Each bin contains approximately 100 flares. Recovery probability is very close to \(1\) for all EDs greater than 100s, showing we recover nearly all flares above this threshold. }
    \label{fig:recov}
\end{figure}

After recovery, the injected flares are separated into 100 bins in log ED. For each bin, we calculate an ED correction by taking the median of the ratios of the true and recovered EDs. We also compute a recovery probability, which measures the percentage of flares in that energy range that were successfully recovered. Figure \ref{fig:recov} shows the recovery probability and ED correction as a function of ED for the K2 target ASASSN-14mz. For all flares with an ED $ >30$ s, the recovery probability is greater than \(0.5\), and it approaches unity around an ED of \(100\) s. There is typically a small ED correction, between \(1\) and \(1.2\) for flare EDs less than \(1000\) s. Table \ref{tab:flares} also includes the recovery probability and corrected ED for each flare.

\subsection{Flare Energy Estimation}
\label{FlareEnergy}
Once we have corrected the ED for each of the K2 and TESS flares, we can compute the bolometric energy for each flare event. For ASASSN-13cm, -14jy, -16dj and -20gu we use parallax data from \textit{Gaia} DR3 to obtain a distance estimate \citep{2018A&A...616A...1G}. 
Due to the presence of nearby background objects, ASASSN-14mz lacks a \textit{Gaia} DR3 parallax. We use the \citet{Schmidt_2019} distance value derived from the \(V-K_s\) magnitude relation \citep{680140829}. Estimated distances for each target are given in Table \ref{tab:targets}. 

Combining the observed stellar flux with the distance estimate gives us the band luminosities for each of our targets. The bolometric energy of a flare is 
\begin{equation}
    E_{\textrm{bol}}=\frac{1}{c_{\textrm{band}}} L_{\textrm{band}} t_{\textrm{ED}}
\end{equation}
where \(t_{\textrm{ED}}\) is the ED, \(L_{\textrm{band}}\) is the quiescent luminosity of the object in the relevant band and \(c_{\textrm{band}}\) is the fraction of the flare bolometric luminosity that lies in the band. Following \citet{Paudel_super} and measurements from \citet{680140843}, we assume the flares to have the spectrum of a 10,000K blackbody. We convolve the blackbody spectrum with the filter response function and integrate, giving us \({c_{\textrm{band}}}=0.21\) for \textit{Kepler} and \({c_{\textrm{band}}}=0.18\) for TESS. This may underestimate the energy in each flare, because it does not include the energy in emission lines that are outside the band \citep{680140843}. Such errors are believed to be small, as the blackbody continuum dominates the flare energy budget in UV/visible wavelengths \citep[e.g.,][]{1991ApJ...378..725H, 2015ApJ...809...79O, Paudel_super}.

\begin{figure}
    \centering
    \includegraphics[width=\linewidth]{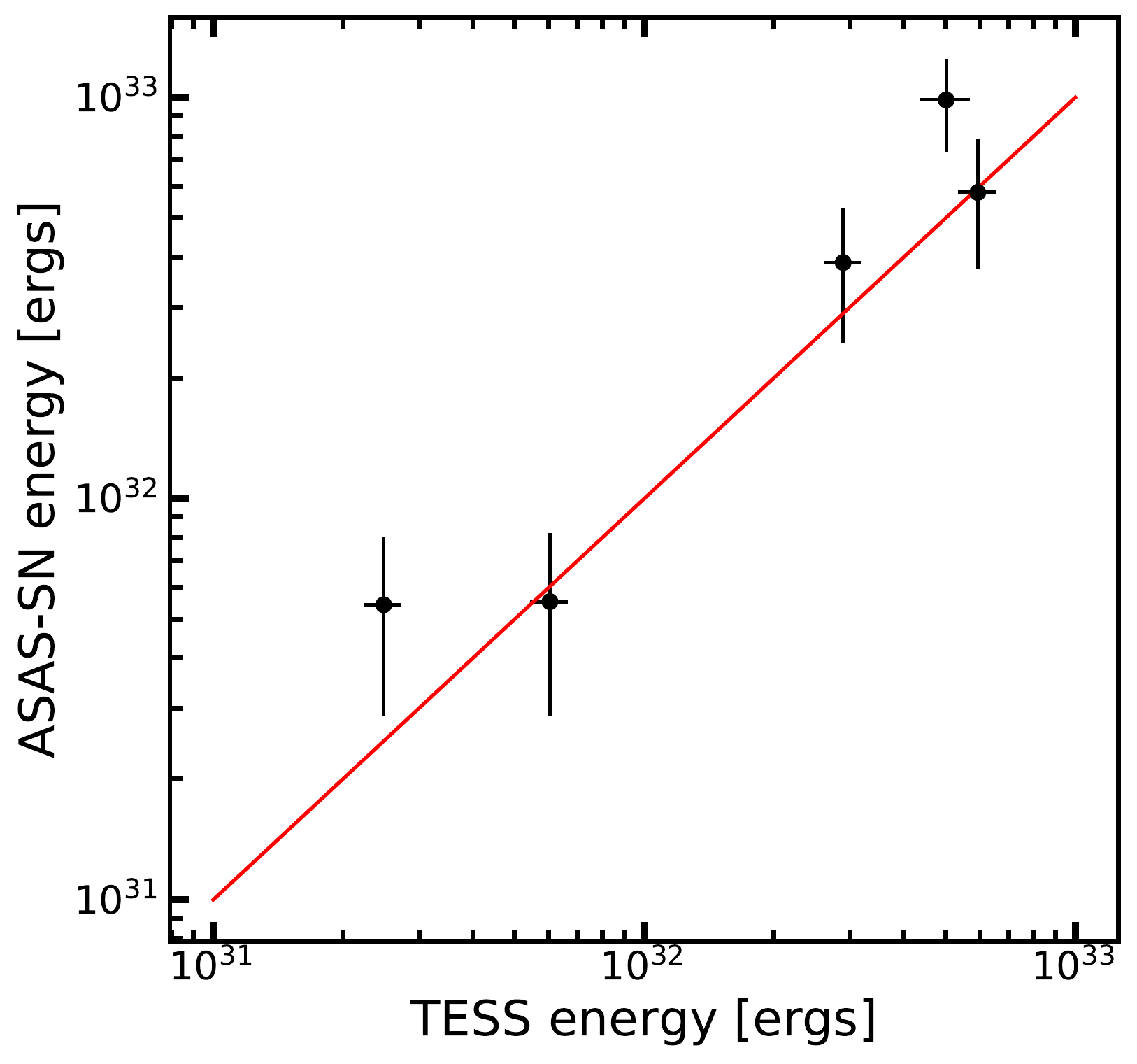}
    \caption{A comparison of the ASAS-SN and TESS flare energy fitting methods for 5 flares on ASASSN-14jy which were observed by both TESS and ASAS-SN. The red line shows where the two energies are equal.}
    \label{fig:overlap}
\end{figure}

To assess the fidelity of the energies estimated from both our fits to the ASAS-SN photometry and the \textit{AltaiPony} procedures used for K2 and TESS, we compare our final flare energies from TESS with the values obtained using our ASAS-SN fitting process for each of the 5 flares that were observed concurrently in TESS and ASAS-SN, which all occurred on the target ASASSN-14jy. As shown in Figure \ref{fig:overlap}, the two estimates are found to be in agreement.

\section{Stellar Rotation}
\label{Gyrochronology}
\label{rotation}
One of the most important questions in M-dwarf dynamics is the relationship between age, rotation rate, and the mechanics of the magnetic dynamo \citep[e.g.][]{Mondrik_2018, gunther}. Measuring rotation rates for our targets is crucial for understanding the superflares that occurred on them.

We compute rotation rates from \textit{Kepler} and TESS lightcurves by tracking the periodic brightening and dimming within a star's lightcurve caused by spots on the stellar surface. We use a combination of three techniques:  Fourier analysis, wavelet analysis, and autocorrelation function (ACF) analysis. \citet{Ceillier_2017} has shown that using these methods together optimizes the recovery of stellar rotation periods .

First, we compute the Lomb-Scargle periodogram for each of our lightcurves using \texttt{lightkurve} \citep{lightkurve}. Lomb-Scargle periodograms build upon Fourier analysis by showing which recovered periods have the strongest power. 
We also use the wavelet transform and the Morlet wavelet in the \texttt{SciPy} library. We compute the wavelet power spectrum and global wavelet power spectrum (GWPS), which is equal to the wavelet power spectrum summed over the full timeseries, once again removing any peaks that correspond to TESS aliases \citep{2020SciPy-NMeth}. Our final rotation period from wavelet analysis is given by the highest peak in the GWPS. 
Finally, we attempt to measure a rotation period using an ACF analysis, which correlates the timeseries data with itself. This method confirms signals found through the Fourier and wavelet analyses while also searching for signals that are not perfectly periodic, sinusoidally shaped, or present throughout the full lightcurve, using the methods outlined in \citet{10.1093/mnras/stt536} and the Python package \texttt{starspot} \citep{starspot} Our final rotation period from ACF analysis is equal to the larger of the first two peaks.

The statistical uncertainties for our rotation measurements are derived from the width of the peak corresponding to the selected period. Along with statistical uncertainty, there is an estimated limit on the precision of any measured rotation period of 10\% due to differential rotation \citep{2014ApJ...780..159E}.  We propagate this systematic uncertainty with the statistical uncertainty from each of our rotation period measurement methods to obtain the final uncertainty on the photometric rotation period during the observations for each star.

For each of our TESS targets, the different methods agreed within the estimated error, indicating a likely detection of rotation. On the other hand, for our K2 target, ASASSN-14mz, there was little agreement between the three methods, with a spread of over 20 days between the estimated periods. Additionally, the peak in the wavelet analysis spectrum was below the cone of influence, which marks where edge effects make results unreliable, indicating that this peak likely does not correspond to a real signal.

\begin{figure*}
    \centering
    \includegraphics[width=\linewidth]{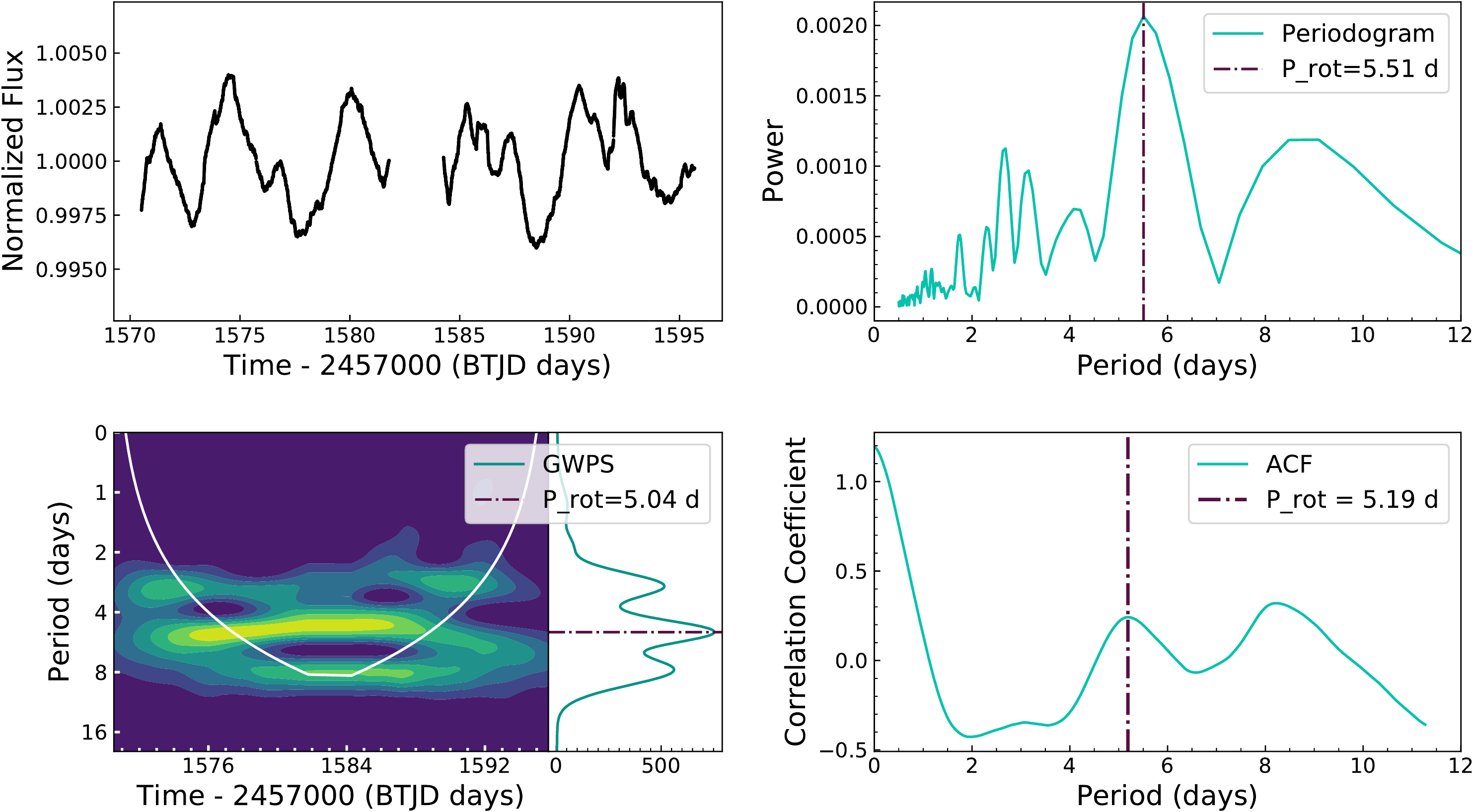}
    \caption{Our rotation analysis for ASASSN-14jy. The full, detrended TESS lightcurve with flares removed is in the top left panel. The bottom left panel shows the results of the wavelet analysis. The white line denotes the cone of influence, below which edge effects make results unreliable.  The top right panel shows the Lomb-Scargle periodogram, while the bottom right panel shows the ACF, where a correlation peak its visible at \(5.19\) days. All three methods find a peak at \(\approx 5.25\) days. The power peaks are clear, and consistent at the 10\% level, evidence of a clear rotation signal present.}
    \label{fig:period}
\end{figure*}

Figure \ref{fig:period} shows an example of our rotation finding methods for the TESS target ASASSN-14jy. All three methods agree to within \(10\)\%, indicating a clear rotation signal at \(\approx 5.25\) days. The rotation periods for each of our targets are given in Table \ref{tab:FFDparams}. For each of the stars that we were able to extract a rotation rate, we found them to be fast rotators, with a period less than 6 days, but more than 1 day. These lie in the range of stars on which studies such as \citet{Mondrik_2018}, \citet{2016ApJ...821L..19N} and \citet{Raetz_2020} have found the most active flaring behaviour. It is also in line with theoretical predictions of a superflaring rate at faster rotation periods.

\section{Flare Distribution}
\label{sec:FlareDist}
\subsection{Estimating Flare Frequency}
\label{ASASFlare}

Because TESS and K2 data are continuous, the observing time is given by the total number of good points multiplied by the observing cadence \citep{680140823}. We therefore compute the cumulative flare frequency for a given flare by counting the number of flares with energy greater than or equal to that flare, and dividing by the total observation time during which that flare could have been observed. We then correct this flare frequency as a function of energy using the recovery probabilities calculated in Section \ref{Injection}.


In order to estimate a FFD for the ASAS-SN flares, we must estimate the observing time during which any given flare would be visible. Because of factors like cloud cover and sky brightness, the minimum detectable flare energy can be different on different nights. Additionally, because ASAS-SN does not stay fixed on a target, we need to estimate the time period before the observation during which the flare could have occurred where the flux level would still be above our sensitivity limit. Due to the challenge of estimating the number of stars on which flares could potentially be detected by ASAS-SN, we do not include the initial flare that triggered the ASAS-SN transient pipeline in our FFDs. However, once we have made our target selection criteria, we are able to search the remaining ASAS-SN data for additional flares, and estimate flare rate as a function of energy. 

To estimate the observing time during which these remaining flares would have been visible, we inject a flare using the standard flare template with energy and \(t_{1/2}\) equal to the value that we recover in Section \ref{ASAS-recov}, again using the flare template from \citet{Davenport_2014}. We test an array of flare start times ranging from 2 hours before the first ASAS-SN data point to 2 minutes after the first ASAS-SN data point, spaced every 30 seconds. A start time is considered recovered if it results in at least 2 data points that exceed the ASAS-SN \(3 \sigma\) detection limit within a given night. We repeat this process for each epoch of ASAS-SN observations, and add together the total recovered time from each night to compute the overall observed time. We then divide the flares by this observed time to get our ASAS-SN flare rates. 

\subsection{Binned and Cumulative FFDs}
\label{sec:FFDSection}
With flare rates for ASAS-SN and K2/TESS data, we can compute a FFD for each of our stars. We compute both a differential, binned FFD, and a cumulative FFD. First, we compute our differential distribution, that allows us to directly correct for our flare detection probability and fit our power law distribution. While previous studies such as \citet{ilan_2018} have computed detection probabilities, they have used them as a cut-off for their cumulative distribution, discarding flares with a detection probability below a certain threshold, as opposed to correcting for them directly. Working with a binned distribution allows us to avoid this problem, and to measure errors in frequency that are not covariant between flares. 

\begin{figure*}
    \centering
    \includegraphics[scale=.55]{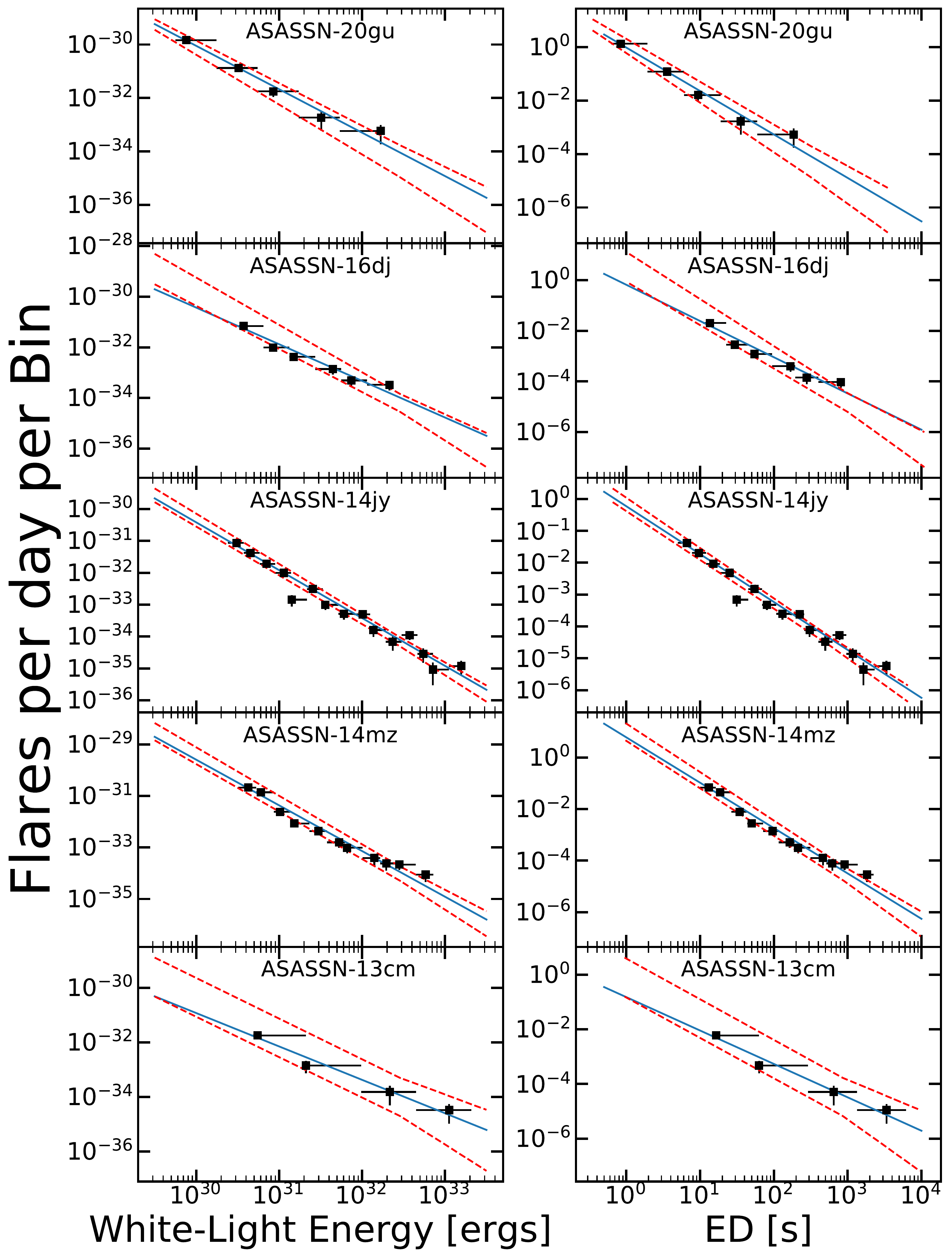}
    \caption{A differential FFD for each of our stars. Plotting frequency against energy per bin allows the best-fit slope to be independent of the number of bins. The x error bars show the full width of each bin. A point is placed at the median ED value in each bin. We fit a power-law to each of the FFDs as described in \ref{sec:FFDSection}. We find each of the slopes to be \(<2\). The red dashed lines show the \(1 \sigma\) error bars for each power law fit.} 
    \label{fig:binned_ffd}
\end{figure*}

For our binned distribution, we separate the flares into bins of equal size in \(\log\) energy, choosing the number of bins to minimise the uncertainty on the final fit. We then divide by the bin width, to normalize for the different bin sizes. We also tested different choices of bin width. The frequency of flares in each bin is given by the number of flares in the bin, divided by the total observing time during which those flares could have been observed. Figure \ref{fig:binned_ffd} shows the differential flare-frequency distributions for each of our targets, in both ED and energy. The horizontal error bars show the full width of the bin in which the frequency was computed and the central point shows the median flare energy in that bin. The error bars on the frequency are derived from the Poisson error added in quadrature to the uncertainty in our recovery probability estimate.

From the earliest studies of flare distributions \citep{GERSHBERG1973, 1976ApJS...30...85L}, the energy distributions of flares have been described by power-laws, capturing the balance between small and large flares. 
To fit our power-law, we take the log of the energy medians and the frequencies, then fit a linear function to the log data using the \texttt{SciPy optimize} least squares optimiser. We choose \(E_\mathrm{{base}}\) to be the median bin location for each star. We report \(E_\mathrm{{base}}\) for each star in Table \ref{tab:FFDparams}. Fitting our data in this manner gives a direct measurement of the slope of the differential distribution, while the slope of the cumulative distribution is given by \(\alpha' = \alpha-1\). We compute power-law fit parameters by bootstrap resampling the flares 1000 times, and report our two-sided error bars as the 16th and 84th percentile of the distributions of \(\alpha\) and \(\beta\) in Table \ref{tab:FFDparams}.

\begin{figure*}
    \centering
    \includegraphics[width=\linewidth]{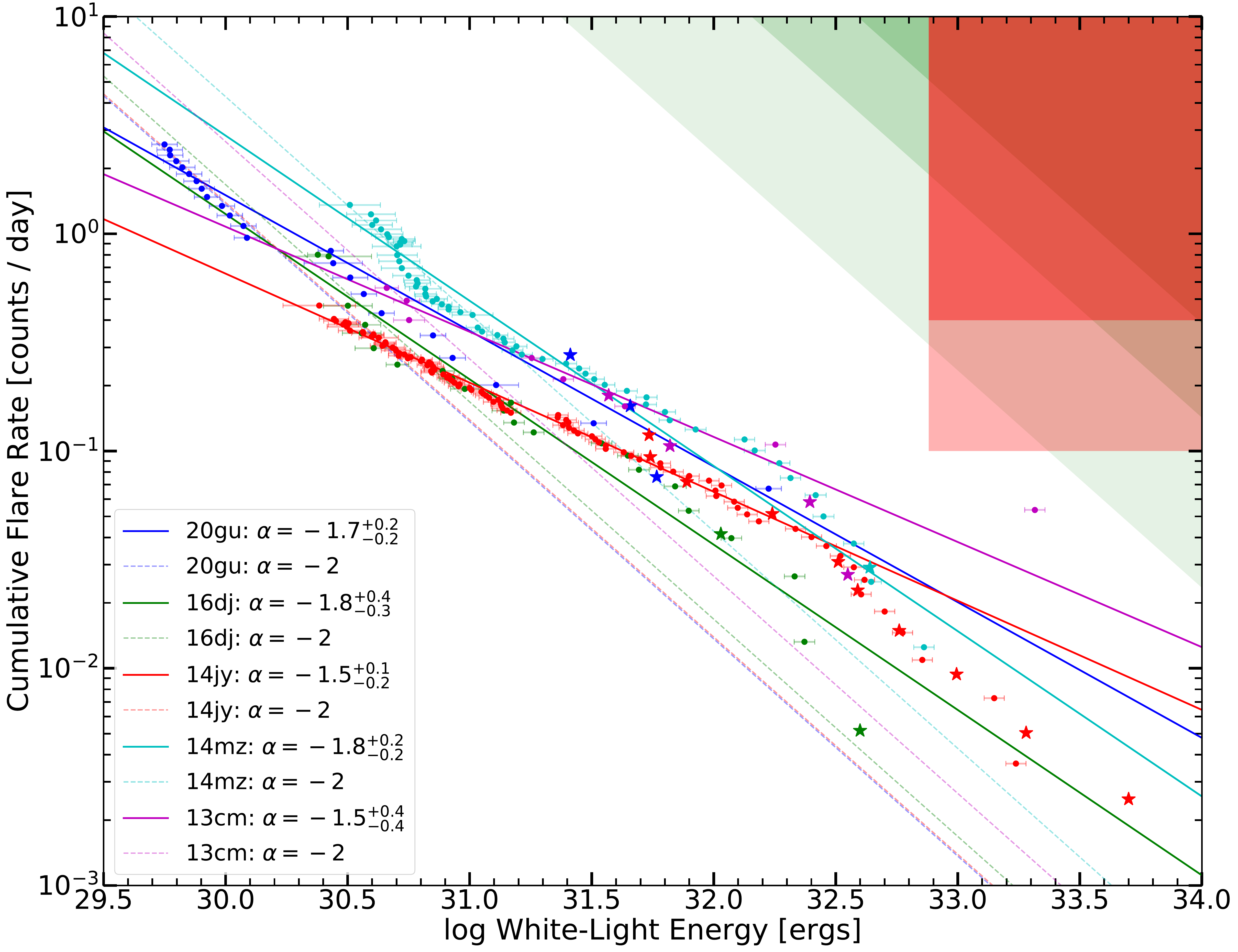}
    \caption{Cumulative FFDs for each of our M-dwarfs. The circles denote each of our TESS/K2 flares, while the stars show our ASAS-SN flares. Error bars on energy come from uncertainty in flux measurements combined with with the uncertainty in the ED correction in our injection and recovery analysis. The solid line fits shown are obtained from the best-fit slope in the differential distribution. The dotted lines show the best fit to a distribution with \(\alpha=2\). The red area denotes the ozone depletion zone, where there is sufficient UV flux in flares to deplete the atmosphere of a Earth-like planet orbiting in the habitable zone of the star. The two color gradients denote the permissive and conservative thresholds given in \citet{gunther}. None of our M-dwarfs' FFDs intersect with this region. The green region denotes the abiogenesis zone form \citet{gunther}, where there is sufficient UV flux in flares to power reactions believed to be necessary for RNA synthesis. The light green denotes the region for M6 stars, the middle green for M4 stars, and the dark green for M3 stars. }
    \label{fig:cum_ffd}
\end{figure*}

Classic, cumulative FFDs for each of our targets are given in Figure \ref{fig:cum_ffd}. For each point, we compute the flare rate for flares with energy greater than or equal to that flare's energy. Due to differences in observation time, the ASAS-SN flares are plotted separately from the K2 and TESS flares. The ASAS-SN flares are indicated with stars, while the K2 and TESS flares are shown using dots. The power-law fits computed from our differential distributions are shown for each of the stars. 

There is significant heterogeneity in the literature both in power-law fitting methods and injection and recovery techniques. Most of the literature fits a power-law directly to the cumulative flare distribution through a maximum likelihood estimator \citep[e.g.][]{ilan_2018}, though some studies, such as \citet{2014ApJ...797..121H}, have fit to a differential distribution. To confirm that our flatter slopes are not the result of fitting method, we also fit each of our cumulative distributions using the maximum likelihood based \texttt{fit\_power\_law} method in \texttt{AltaiPony}. We find that each cumulative FFD power-law fit still has \(\alpha<2\). Our use of an injection and recovery method is unlikely to be the source of our unusually flat FFD slopes, as the process makes the recovered power-law slope steeper, due to the correction increasing flare rates at lower energies.

\begin{deluxetable*}{ccccccccc}[htbp!]
\tablecaption{FFD Parameters}
\tablehead{
\colhead{Target} &
\colhead{Rotation Period} &
\colhead{Number of Flares} &
\colhead{\(\alpha\)} &
\colhead{\(\beta\)} & \colhead{\(E_\mathrm{{base}}\)}
\\
&
 \colhead{(days)} &
\colhead{(TESS/K2, ASAS-SN V-band, $g$-band)} &
&
\colhead{($10^{-32}$ flares/day)} & (erg)}
 
\startdata
   ASASSN-13cm & $1.21 \pm 0.01$ & 8, 1, 4 & $-1.31^{+0.42}_{-0.43}$ & $1.03_{ -0.47}^{+0.5}$ & $1.1*10^{31}$  \\
ASASSN-14jy & $5.25 \pm 0.2$ & 97, 4, 8 & $ -1.55_{ -0.12}^{+0.11}$& $1.23_{ -0.15}^{+0.22}$ & $8.1*10^{31}$ \\ 
ASASSN-14mz & N/A & 58, 2, 0 & $-1.82_{ -0.24}^{+0.17}$ & $4.91_{ -0.65}^{+0.81}$ & $5.3*10^{31}$   \\ 
ASASSN-16dj & $1.21 \pm 0.01$ & 21, 1, 2 & $-1.76_{ -0.3}^{+0.35}$ & $1.90_{ -0.59}^{+0.81}$ & $4.4*10^{31}$ \\ 
ASASSN-20gu & \(5.24 \pm 0.3\) & 23, 0, 5 &  $  -1.71_{ -0.27}^{+0.23}$ & $1.95_{ -0.81}^{+1.35}$ & $8.5*10^{30}$ \\ 
\enddata 
\tablecomments{The essential FFD parameters for each of our targets. The rotation period is given as the average of our 3 methods described in Section \ref{rotation} with the full range of error reported. \(\alpha\) and \(\beta\) are the parameters for the best power law fit to our FFD. An explanation of the fitting process is given in Section \ref{ASASFlare}.}
\label{tab:FFDparams}
\end{deluxetable*}
 
\section{Discussion and Conclusions}

\label{discussion}

We analyze the flare distributions of 5 M-dwarfs which were observed undergoing a superflare in ASAS-SN and were also observed at short cadence by K2 or TESS for at least one sector. We use the \texttt{FINDflare} algorithm to identify 207 flares in the detrended K2 and TESS data. We also identify 19 flares in the ASAS-SN $g$-band data. We run injection and recovery tests for each of our stars, and derive ED corrections and recovery probabilities for each of the flares in our sample. We also measure rotation rates for 4 of our stars, and find each of them to be fast-rotating, with \(P<10\) days. This agrees with the findings of \citet{Mondrik_2018} that superflares are more likely to occur on faster rotating, more magnetically active stars.

We present the cumulative FFDs for each star, as well as a binned FFD, which allows us to directly correct for recovery probability, and to remove covariances between the frequency errors of points. Fitting a power-law to each of the flare distributions, we find that all of them have slopes flatter than \(\alpha = 2 \). These results on superflaring M-dwarfs disagree with  \citet{ilan_2018} and \citet{2019ApJ...881....9H} who argue that most M-dwarfs, regardless of properties, have \(\alpha \approx 2\).

The FFD slope \(\alpha\) plays an important role in the physics of stellar coronae. \citet{1990A&A...232...83D} first proposed flares as a possible explanation for high levels of coronal heating. Then, \citet{2004A&ARv..12...71G} found that sufficient energy required to power coronal heating could be provided by many low-energy microflares. If the distribution of flares were sufficiently steep, with \(\alpha>2\), then low-energy flares could fully heat the corona. \citet{ilan_2018} found that a sample of stars in open clusters observed using K2 short cadence data all had \(2 \leq \alpha \leq 2.4\), suggesting that the coronae on young M-dwarfs could be primarily flare-powered. Conversely, each of our targets have a flare-frequency distribution with \(\alpha \leq 2\), suggesting that weak flares are unlikely to be the primary source of a coronal heating mechanism on these stars.

In addition to the impact of flares on stellar properties, large flares may also impact the atmospheres of exoplanets orbiting M-dwarfs. The largest superflares on M-dwarfs are often orders of magnitude larger than the largest Solar flares recorded, and the habitable zone of M-dwarfs are much smaller than 1 AU. So possible life-bearing planets orbiting M-dwarfs could experience temporary UV fluxes 4 or more orders of magnitude larger than those which regularly strike the Earth. Estimates of the rate of superflaring with \(E\geq 10^{34}\) erg needed to fully deplete the ozone layer of an Earth-like planet in the habitable zone of a mid M-dwarf range from  approximately 0.1 $\text{flares day}^{-1}$ to  0.4 $\text{flares day}^{-1}$ \citep{Tilley2019, gunther}. Figure \ref{fig:cum_ffd} shades the region leading to ozone depletion in red. Each of our targets falls below the region, even though ASASSN-13cm has a relatively high superflare frequency of \(0.064\) flares/day. So, even though our targets were selected for superflaring activity, and have significantly flatter FFDs, ozone depletion from direct UV emission is unlikely. This agrees with the results of \citet{gunther}, who found only 15 M-dwarfs in the ozone depletion region out of a sample of 1228 flaring stars.

The impact of stellar flares on nearby exoplanet habitability is not always negative. Due to their low effective temperatures, quiescent M dwarfs provide significantly lower levels of UV flux to planets orbiting them than solar type stars \citep{2018SciA....4.3302R}. This may hinder the development of Earth-like life on these exoplanets, as laboratory studies have found that sufficiently high UV fluxes (about those of a K5 dwarf with \(T_{\mathrm{eff}}=4400\mathrm{K}\)) are required to enable the prebiotic chemical reactions that produce important precursors to biological molecules like ribonucleic acid (RNA; \citealt{2018SciA....4.3302R}). Flares can help mitigate this low steady-state UV flux. Following the methods of \citet{2018SciA....4.3302R} and \citet{gunther}, we shade the region of FFD phase space in which abiogenesis is possible. None of our targets intersect with this zone at energies below \(10^{35}\) erg. Our results indicate that even active, superflaring stars are unlikely to produce the UV fluxes needed for the synthesis of these molecules, in agreement with \citet{gunther}. 

Data from surveys like ASAS-SN in combination with short-cadence K2 and TESS data and effective automated flare-finding techniques have made a broader understanding of flaring on low-mass dwarfs possible. These new data enable studies of the statistics of flare distributions across a much larger sample of stars, which can be particularly helpful in deciphering the underlying physical mechanisms behind flares. As TESS puts more stars on a 20s cadence in its extended mission, we will be able to probe even lower energy flares, allowing us to better understand what typical flaring M-dwarfs look like, and how these flares impact their evolution and the evolution of the exoplanets that orbit them. 

\section*{Acknowledgements}

We thank Nicholas Saunders and Ekaterina Ilin for helpful discussions, and Michael Tucker for his helpful comments on the manuscript. 

JZ acknowledges support from Research Experience for Undergraduate programme  at the Institute for Astronomy, University of Hawaii-Manoa funded through  NSF grant 6104374.
JZ would also like to thank the Institute for Astronomy for their kind hospitality during the course of this project.

We thank the Las Cumbres Observatory and its staff for its continuing support of the ASAS-SN project. ASAS-SN is supported by the Gordon and Betty Moore Foundation through  grant GBMF5490 to the Ohio State University, and NSF grants AST-1515927 and AST-1908570. Development of ASAS-SN has been supported by NSF grant AST-0908816, the Mt. Cuba Astronomical Foundation, the Centre  for Cosmology and AstroParticle Physics at the Ohio State University, the Chinese Academy of Sciences South America Centre  for Astronomy (CAS- SACA), and the Villum Foundation. 

JTH is supported by NASA award 80NSSC21K0136. BJS, CSK, and KZS are supported by NSF grant AST-1907570. BJS is also supported by NASA grant 80NSSC19K1717 and NSF grants AST-1920392 and AST-1911074. CSK and KZS are supported by NSF grant AST-181440. Support for JLP is provided in part by FONDECYT through  the grant 1151445 and by the Ministry of Economy, Development, and Tourism's Millennium Science Initiative through  grant IC120009, awarded to The Millennium Institute of Astrophysics, MAS. TAT is supported in part by Scialog Scholar grant 24215 from the Research Corporation.

Support for TW-SH was provided by NASA through the NASA Hubble Fellowship grant HST-HF2-51458.001-A awarded by the Space Telescope Science Institute, which is operated by the Association of Universities for Research in Astronomy, Inc., for NASA, under contract NAS5-26555.

This work is based on observations made by ASAS-SN. We wish to extend our special thanks to those of Hawaiian ancestry on whose sacred mountain of Maunakea we are privileged to be guests. Without their generous hospitality, the observations presented herein would not have been possible.

\bibliography{references.bib}{}

\begin{thebibliography}{}
\expandafter\ifx\csname natexlab\endcsname\relax\def\natexlab#1{#1}\fi
\providecommand{\url}[1]{\href{#1}{#1}}
\providecommand{\dodoi}[1]{doi:~\href{http://doi.org/#1}{\nolinkurl{#1}}}
\providecommand{\doeprint}[1]{\href{http://ascl.net/#1}{\nolinkurl{http://ascl.net/#1}}}
\providecommand{\doarXiv}[1]{\href{https://arxiv.org/abs/#1}{\nolinkurl{https://arxiv.org/abs/#1}}}

\bibitem[{{Aigrain} {et~al.}(2016){Aigrain}, {Parviainen}, \& {Pope}}]{Aigrain}
{Aigrain}, S., {Parviainen}, H., \& {Pope}, B.~J.~S. 2016, \mnras, 459, 2408,
  \dodoi{10.1093/mnras/stw706}

\bibitem[{{Alard}(2000)}]{alard00}
{Alard}, C. 2000, \aaps, 144, 363, \dodoi{10.1051/aas:2000214}

\bibitem[{{Alard} \& {Lupton}(1998)}]{alard98}
{Alard}, C., \& {Lupton}, R.~H. 1998, \apj, 503, 325, \dodoi{10.1086/305984}

\bibitem[{Angus(2021)}]{starspot}
Angus, R. 2021, {RuthAngus/starspot: code for measuring stellar rotation
  periods}, v0.2,  Zenodo, \dodoi{10.5281/zenodo.4613887}

\bibitem[{{Borucki} {et~al.}(2010){Borucki}, {Koch}, {Basri}, {Batalha},
  {Brown}, {Caldwell}, {Caldwell}, {Christensen-Dalsgaard}, {Cochran},
  {DeVore}, {Dunham}, {Dupree}, {Gautier}, {Geary}, {Gilliland}, {Gould},
  {Howell}, {Jenkins}, {Kondo}, {Latham}, {Marcy}, {Meibom}, {Kjeldsen},
  {Lissauer}, {Monet}, {Morrison}, {Sasselov}, {Tarter}, {Boss}, {Brownlee},
  {Owen}, {Buzasi}, {Charbonneau}, {Doyle}, {Fortney}, {Ford}, {Holman},
  {Seager}, {Steffen}, {Welsh}, {Rowe}, {Anderson}, {Buchhave}, {Ciardi},
  {Walkowicz}, {Sherry}, {Horch}, {Isaacson}, {Everett}, {Fischer}, {Torres},
  {Johnson}, {Endl}, {MacQueen}, {Bryson}, {Dotson}, {Haas}, {Kolodziejczak},
  {Van Cleve}, {Chandrasekaran}, {Twicken}, {Quintana}, {Clarke}, {Allen},
  {Li}, {Wu}, {Tenenbaum}, {Verner}, {Bruhweiler}, {Barnes}, \&
  {Prsa}}]{Borucki}
{Borucki}, W.~J., {Koch}, D., {Basri}, G., {et~al.} 2010, Science, 327, 977,
  \dodoi{10.1126/science.1185402}

\bibitem[{Browning(2008{\natexlab{a}})}]{Browning_2008}
Browning, M.~K. 2008{\natexlab{a}}, The Astrophysical Journal, 676,
  1262–1280, \dodoi{10.1086/527432}

\bibitem[{Browning(2008{\natexlab{b}})}]{680140810}
---. 2008{\natexlab{b}}, ApJ, 676, 1262.
\newblock \url{http://stacks.iop.org/0004-637X/676/i=2/a=1262}

\bibitem[{{Ceillier} {et~al.}(2017){Ceillier}, {Tayar}, {Mathur}, {Salabert},
  {Garc{\'{\i}}a}, {Stello}, {Pinsonneault}, {van Saders}, {Beck}, \&
  {Bloemen}}]{Ceillier_2017}
{Ceillier}, T., {Tayar}, J., {Mathur}, S., {et~al.} 2017, \aap, 605, A111,
  \dodoi{10.1051/0004-6361/201629884}

\bibitem[{{Chang} {et~al.}(2015){Chang}, {Byun}, \&
  {Hartman}}]{2015ApJ...814...35C}
{Chang}, S.~W., {Byun}, Y.~I., \& {Hartman}, J.~D. 2015, \apj, 814, 35,
  \dodoi{10.1088/0004-637X/814/1/35}

\bibitem[{{Davenport}(2016)}]{2016ApJ...829...23D}
{Davenport}, J. R.~A. 2016, \apj, 829, 23, \dodoi{10.3847/0004-637X/829/1/23}

\bibitem[{{Davenport} {et~al.}(2012){Davenport}, {Becker}, {Kowalski},
  {Hawley}, {Schmidt}, {Hilton}, {Sesar}, \& {Cutri}}]{2012ApJ...748...58D}
{Davenport}, J. R.~A., {Becker}, A.~C., {Kowalski}, A.~F., {et~al.} 2012, \apj,
  748, 58, \dodoi{10.1088/0004-637X/748/1/58}

\bibitem[{{Davenport} {et~al.}(2019){Davenport}, {Covey}, {Clarke}, {Boeck},
  {Cornet}, \& {Hawley}}]{2019ApJ...871..241D}
{Davenport}, J. R.~A., {Covey}, K.~R., {Clarke}, R.~W., {et~al.} 2019, \apj,
  871, 241, \dodoi{10.3847/1538-4357/aafb76}

\bibitem[{{Davenport} {et~al.}(2020){Davenport}, {Tovar Mendoza}, \&
  {Hawley}}]{2020arXiv200510281D}
{Davenport}, J. R.~A., {Tovar Mendoza}, G., \& {Hawley}, S.~L. 2020, arXiv
  e-prints, arXiv:2005.10281.
\newblock \doarXiv{2005.10281}

\bibitem[{Davenport {et~al.}(2014)Davenport, Hawley, Hebb, Wisniewski,
  Kowalski, Johnson, Malatesta, Peraza, Keil, Silverberg, \&
  et~al.}]{Davenport_2014}
Davenport, J. R.~A., Hawley, S.~L., Hebb, L., {et~al.} 2014, The Astrophysical
  Journal, 797, 122, \dodoi{10.1088/0004-637x/797/2/122}

\bibitem[{{Doyle} {et~al.}(1990){Doyle}, {Butler}, {van den Oord}, \&
  {Kiang}}]{1990A&A...232...83D}
{Doyle}, J.~G., {Butler}, C.~J., {van den Oord}, G.~H.~J., \& {Kiang}, T. 1990,
  \aap, 232, 83

\bibitem[{{Dressing} \& {Charbonneau}(2015)}]{2015ApJ...807...45D}
{Dressing}, C.~D., \& {Charbonneau}, D. 2015, \apj, 807, 45,
  \dodoi{10.1088/0004-637X/807/1/45}

\bibitem[{{Epstein} \& {Pinsonneault}(2014)}]{2014ApJ...780..159E}
{Epstein}, C.~R., \& {Pinsonneault}, M.~H. 2014, \apj, 780, 159,
  \dodoi{10.1088/0004-637X/780/2/159}

\bibitem[{Feinstein {et~al.}(2020)Feinstein, Montet, Ansdell, Nord, Bean,
  Günther, Gully-Santiago, \& Schlieder}]{Feinstein_2020}
Feinstein, A.~D., Montet, B.~T., Ansdell, M., {et~al.} 2020, The Astronomical
  Journal, 160, 219, \dodoi{10.3847/1538-3881/abac0a}

\bibitem[{{Gaia Collaboration} {et~al.}(2018){Gaia Collaboration}, {Brown},
  {Vallenari}, {Prusti}, {de Bruijne}, {Babusiaux}, {Bailer-Jones}, {Biermann},
  {Evans}, {Eyer}, {Jansen}, {Jordi}, {Klioner}, {Lammers}, {Lindegren},
  {Luri}, {Mignard}, {Panem}, {Pourbaix}, {Randich}, {Sartoretti}, {Siddiqui},
  {Soubiran}, {van Leeuwen}, {Walton}, {Arenou}, {Bastian}, {Cropper},
  {Drimmel}, {Katz}, {Lattanzi}, {Bakker}, {Cacciari}, {Casta{\~n}eda},
  {Chaoul}, {Cheek}, {De Angeli}, {Fabricius}, {Guerra}, {Holl}, {Masana},
  {Messineo}, {Mowlavi}, {Nienartowicz}, {Panuzzo}, {Portell}, {Riello},
  {Seabroke}, {Tanga}, {Th{\'e}venin}, {Gracia-Abril}, {Comoretto},
  {Garcia-Reinaldos}, {Teyssier}, {Altmann}, {Andrae}, {Audard},
  {Bellas-Velidis}, {Benson}, {Berthier}, {Blomme}, {Burgess}, {Busso},
  {Carry}, {Cellino}, {Clementini}, {Clotet}, {Creevey}, {Davidson}, {De
  Ridder}, {Delchambre}, {Dell'Oro}, {Ducourant},
  {Fern{\'a}ndez-Hern{\'a}ndez}, {Fouesneau}, {Fr{\'e}mat}, {Galluccio},
  {Garc{\'\i}a-Torres}, {Gonz{\'a}lez-N{\'u}{\~n}ez}, {Gonz{\'a}lez-Vidal},
  {Gosset}, {Guy}, {Halbwachs}, {Hambly}, {Harrison}, {Hern{\'a}ndez},
  {Hestroffer}, {Hodgkin}, {Hutton}, {Jasniewicz}, {Jean-Antoine-Piccolo},
  {Jordan}, {Korn}, {Krone-Martins}, {Lanzafame}, {Lebzelter}, {L{\"o}ffler},
  {Manteiga}, {Marrese}, {Mart{\'\i}n-Fleitas}, {Moitinho}, {Mora}, {Muinonen},
  {Osinde}, {Pancino}, {Pauwels}, {Petit}, {Recio-Blanco}, {Richards},
  {Rimoldini}, {Robin}, {Sarro}, {Siopis}, {Smith}, {Sozzetti}, {S{\"u}veges},
  {Torra}, {van Reeven}, {Abbas}, {Abreu Aramburu}, {Accart}, {Aerts},
  {Altavilla}, {{\'A}lvarez}, {Alvarez}, {Alves}, {Anderson}, {Andrei},
  {Anglada Varela}, {Antiche}, {Antoja}, {Arcay}, {Astraatmadja}, {Bach},
  {Baker}, {Balaguer-N{\'u}{\~n}ez}, {Balm}, {Barache}, {Barata}, {Barbato},
  {Barblan}, {Barklem}, {Barrado}, {Barros}, {Barstow}, {Bartholom{\'e}
  Mu{\~n}oz}, {Bassilana}, {Becciani}, {Bellazzini}, {Berihuete}, {Bertone},
  {Bianchi}, {Bienaym{\'e}}, {Blanco-Cuaresma}, {Boch}, {Boeche}, {Bombrun},
  {Borrachero}, {Bossini}, {Bouquillon}, {Bourda}, {Bragaglia}, {Bramante},
  {Breddels}, {Bressan}, {Brouillet}, {Br{\"u}semeister}, {Brugaletta},
  {Bucciarelli}, {Burlacu}, {Busonero}, {Butkevich}, {Buzzi}, {Caffau},
  {Cancelliere}, {Cannizzaro}, {Cantat-Gaudin}, {Carballo}, {Carlucci},
  {Carrasco}, {Casamiquela}, {Castellani}, {Castro-Ginard}, {Charlot},
  {Chemin}, {Chiavassa}, {Cocozza}, {Costigan}, {Cowell}, {Crifo}, {Crosta},
  {Crowley}, {Cuypers}, {Dafonte}, {Damerdji}, {Dapergolas}, {David}, {David},
  {de Laverny}, {De Luise}, {De March}, {de Martino}, {de Souza}, {de Torres},
  {Debosscher}, {del Pozo}, {Delbo}, {Delgado}, {Delgado}, {Di Matteo},
  {Diakite}, {Diener}, {Distefano}, {Dolding}, {Drazinos}, {Dur{\'a}n},
  {Edvardsson}, {Enke}, {Eriksson}, {Esquej}, {Eynard Bontemps}, {Fabre},
  {Fabrizio}, {Faigler}, {Falc{\~a}o}, {Farr{\`a}s Casas}, {Federici},
  {Fedorets}, {Fernique}, {Figueras}, {Filippi}, {Findeisen}, {Fonti},
  {Fraile}, {Fraser}, {Fr{\'e}zouls}, {Gai}, {Galleti}, {Garabato},
  {Garc{\'\i}a-Sedano}, {Garofalo}, {Garralda}, {Gavel}, {Gavras}, {Gerssen},
  {Geyer}, {Giacobbe}, {Gilmore}, {Girona}, {Giuffrida}, {Glass}, {Gomes},
  {Granvik}, {Gueguen}, {Guerrier}, {Guiraud}, {Guti{\'e}rrez-S{\'a}nchez},
  {Haigron}, {Hatzidimitriou}, {Hauser}, {Haywood}, {Heiter}, {Helmi}, {Heu},
  {Hilger}, {Hobbs}, {Hofmann}, {Holland}, {Huckle}, {Hypki}, {Icardi},
  {Jan{\ss}en}, {Jevardat de Fombelle}, {Jonker}, {Juh{\'a}sz}, {Julbe},
  {Karampelas}, {Kewley}, {Klar}, {Kochoska}, {Kohley}, {Kolenberg},
  {Kontizas}, {Kontizas}, {Koposov}, {Kordopatis}, {Kostrzewa-Rutkowska},
  {Koubsky}, {Lambert}, {Lanza}, {Lasne}, {Lavigne}, {Le Fustec}, {Le
  Poncin-Lafitte}, {Lebreton}, {Leccia}, {Leclerc}, {Lecoeur-Taibi},
  {Lenhardt}, {Leroux}, {Liao}, {Licata}, {Lindstr{\o}m}, {Lister}, {Livanou},
  {Lobel}, {L{\'o}pez}, {Managau}, {Mann}, {Mantelet}, {Marchal}, {Marchant},
  {Marconi}, {Marinoni}, {Marschalk{\'o}}, {Marshall}, {Martino}, {Marton},
  {Mary}, {Massari}, {Matijevi{\v{c}}}, {Mazeh}, {McMillan}, {Messina},
  {Michalik}, {Millar}, {Molina}, {Molinaro}, {Moln{\'a}r}, {Montegriffo},
  {Mor}, {Morbidelli}, {Morel}, {Morris}, {Mulone}, {Muraveva}, {Musella},
  {Nelemans}, {Nicastro}, {Noval}, {O'Mullane}, {Ord{\'e}novic},
  {Ord{\'o}{\~n}ez-Blanco}, {Osborne}, {Pagani}, {Pagano}, {Pailler},
  {Palacin}, {Palaversa}, {Panahi}, {Pawlak}, {Piersimoni}, {Pineau}, {Plachy},
  {Plum}, {Poggio}, {Poujoulet}, {Pr{\v{s}}a}, {Pulone}, {Racero}, {Ragaini},
  {Rambaux}, {Ramos-Lerate}, {Regibo}, {Reyl{\'e}}, {Riclet}, {Ripepi}, {Riva},
  {Rivard}, {Rixon}, {Roegiers}, {Roelens}, {Romero-G{\'o}mez}, {Rowell},
  {Royer}, {Ruiz-Dern}, {Sadowski}, {Sagrist{\`a} Sell{\'e}s}, {Sahlmann},
  {Salgado}, {Salguero}, {Sanna}, {Santana-Ros}, {Sarasso}, {Savietto},
  {Schultheis}, {Sciacca}, {Segol}, {Segovia}, {S{\'e}gransan}, {Shih},
  {Siltala}, {Silva}, {Smart}, {Smith}, {Solano}, {Solitro}, {Sordo}, {Soria
  Nieto}, {Souchay}, {Spagna}, {Spoto}, {Stampa}, {Steele},
  {Steidelm{\"u}ller}, {Stephenson}, {Stoev}, {Suess}, {Surdej}, {Szabados},
  {Szegedi-Elek}, {Tapiador}, {Taris}, {Tauran}, {Taylor}, {Teixeira},
  {Terrett}, {Teyssand ier}, {Thuillot}, {Titarenko}, {Torra Clotet}, {Turon},
  {Ulla}, {Utrilla}, {Uzzi}, {Vaillant}, {Valentini}, {Valette}, {van Elteren},
  {Van Hemelryck}, {van Leeuwen}, {Vaschetto}, {Vecchiato}, {Veljanoski},
  {Viala}, {Vicente}, {Vogt}, {von Essen}, {Voss}, {Votruba}, {Voutsinas},
  {Walmsley}, {Weiler}, {Wertz}, {Wevers}, {Wyrzykowski}, {Yoldas},
  {{\v{Z}}erjal}, {Ziaeepour}, {Zorec}, {Zschocke}, {Zucker}, {Zurbach}, \&
  {Zwitter}}]{2018A&A...616A...1G}
{Gaia Collaboration}, {Brown}, A.~G.~A., {Vallenari}, A., {et~al.} 2018, \aap,
  616, A1, \dodoi{10.1051/0004-6361/201833051}

\bibitem[{Gershburg \& Shakahovskaya(1973)}]{GERSHBERG1973}
Gershburg, R.~E., \& Shakahovskaya, N.~I. 1973, Nature Physical Science, 242,
  85, \dodoi{10.1038/physci242085a0}

\bibitem[{Gizis {et~al.}(2013)Gizis, Burgasser, \& et~al}]{680140823}
Gizis, J.~E., Burgasser, A.~J., \& et~al, E.~B. 2013, ApJ, 779, 172.
\newblock \url{http://stacks.iop.org/0004-637X/779/i=2/a=172}

\bibitem[{Gizis {et~al.}(2017)Gizis, Paudel, Schmidt, Williams, \&
  Burgasser}]{Gizis_2017}
Gizis, J.~E., Paudel, R.~R., Schmidt, S.~J., Williams, P. K.~G., \& Burgasser,
  A.~J. 2017, The Astrophysical Journal, 838, 22,
  \dodoi{10.3847/1538-4357/aa6197}

\bibitem[{Grootel {et~al.}(2018)Grootel, Fernandes, Gillon, Jehin, Manfroid,
  Scuflaire, Burgasser, Barkaoui, Benkhaldoun, Burdanov, Delrez, Demory,
  de~Wit, Queloz, \& Triaud}]{Grootel_2018}
Grootel, V.~V., Fernandes, C.~S., Gillon, M., {et~al.} 2018, The Astrophysical
  Journal, 853, 30, \dodoi{10.3847/1538-4357/aaa023}

\bibitem[{{G{\"u}del}(2004)}]{2004A&ARv..12...71G}
{G{\"u}del}, M. 2004, \aapr, 12, 71, \dodoi{10.1007/s00159-004-0023-2}

\bibitem[{{G{\"u}nther} {et~al.}(2020){G{\"u}nther}, {Zhan}, {Seager},
  {Rimmer}, {Ranjan}, {Stassun}, {Oelkers}, {Daylan}, {Newton}, {Kristiansen},
  {Olah}, {Gillen}, {Rappaport}, {Ricker}, {Vanderspek}, {Latham}, {Winn},
  {Jenkins}, {Glidden}, {Fausnaugh}, {Levine}, {Dittmann}, {Quinn},
  {Krishnamurthy}, \& {Ting}}]{gunther}
{G{\"u}nther}, M.~N., {Zhan}, Z., {Seager}, S., {et~al.} 2020, \aj, 159, 60,
  \dodoi{10.3847/1538-3881/ab5d3a}

\bibitem[{Hawley {et~al.}(2014)Hawley, A., \& et~al}]{680140826}
Hawley, S.~L., A., D. J.~R., \& et~al, K. A.~F. 2014, ApJ, 797, 121.
\newblock \url{http://stacks.iop.org/0004-637X/797/i=2/a=121}

\bibitem[{{Hawley} {et~al.}(2014){Hawley}, {Davenport}, {Kowalski},
  {Wisniewski}, {Hebb}, {Deitrick}, \& {Hilton}}]{2014ApJ...797..121H}
{Hawley}, S.~L., {Davenport}, J. R.~A., {Kowalski}, A.~F., {et~al.} 2014, \apj,
  797, 121, \dodoi{10.1088/0004-637X/797/2/121}

\bibitem[{{Hawley} \& {Pettersen}(1991)}]{1991ApJ...378..725H}
{Hawley}, S.~L., \& {Pettersen}, B.~R. 1991, \apj, 378, 725,
  \dodoi{10.1086/170474}

\bibitem[{{Henden} {et~al.}(2015){Henden}, {Levine}, {Terrell}, \&
  {Welch}}]{henden15}
{Henden}, A.~A., {Levine}, S., {Terrell}, D., \& {Welch}, D.~L. 2015, in
  American Astronomical Society Meeting Abstracts, Vol. 225, American
  Astronomical Society Meeting Abstracts, 336.16

\bibitem[{Henry {et~al.}(2004)Henry, James, \& et~al}]{680140829}
Henry, James, \& et~al, K. 2004, AJ, 128, 2460.
\newblock \url{http://stacks.iop.org/1538-3881/128/i=5/a=2460}

\bibitem[{{Hilton}(2011{\natexlab{a}})}]{2011PhDT.......144H}
{Hilton}, E.~J. 2011{\natexlab{a}}, PhD thesis, University of Washington

\bibitem[{{Hilton}(2011{\natexlab{b}})}]{680140898}
---. 2011{\natexlab{b}}, PhD thesis, University of Washington

\bibitem[{Hilton {et~al.}(2010)Hilton, West, Hawley, \& Kowalski}]{680140831}
Hilton, E.~J., West, A.~A., Hawley, S.~L., \& Kowalski, A.~F. 2010, AJ, 140,
  1402.
\newblock \url{http://stacks.iop.org/1538-3881/140/i=5/a=1402}

\bibitem[{{Howard} {et~al.}(2019){Howard}, {Corbett}, {Law}, {Ratzloff},
  {Glazier}, {Fors}, {del Ser}, \& {Haislip}}]{2019ApJ...881....9H}
{Howard}, W.~S., {Corbett}, H., {Law}, N.~M., {et~al.} 2019, \apj, 881, 9,
  \dodoi{10.3847/1538-4357/ab2767}

\bibitem[{Howell {et~al.}(2014)Howell, C., \& et~al}]{680140837}
Howell, C., S., \& et~al, H.~M. 2014, PASP, 126, 398.
\newblock \url{http://stacks.iop.org/1538-3873/126/i=938/a=398}

\bibitem[{{Ilin} {et~al.}(2020){Ilin}, {Schmidt}, {Poppenh{\"a}ger},
  {Davenport}, {Kristiansen}, \& {Omohundro}}]{2020arXiv201005576I}
{Ilin}, E., {Schmidt}, S.~J., {Poppenh{\"a}ger}, K., {et~al.} 2020, arXiv
  e-prints, arXiv:2010.05576.
\newblock \doarXiv{2010.05576}

\bibitem[{Ilin {et~al.}(2019)Ilin, {Schmidt, Sarah J.}, {Davenport, James R.
  A.}, \& {Strassmeier, Klaus G.}}]{ilan_2018}
Ilin, E., {Schmidt, Sarah J.}, {Davenport, James R. A.}, \& {Strassmeier, Klaus
  G.} 2019, A\&A, 622, A133, \dodoi{10.1051/0004-6361/201834400}

\bibitem[{Khodachenko {et~al.}(2007)Khodachenko, I, \& et~al}]{680140839}
Khodachenko, M.~L., I, R., \& et~al, L.~H. 2007, AsBio, 7, 167.
\newblock \url{http://dx.doi.org/10.1089/ast.2006.0127}

\bibitem[{Kochanek {et~al.}(2017)Kochanek, J., \& et~al}]{680140840}
Kochanek, C.~S., J., S.~B., \& et~al, S. K.~Z. 2017, PASP, 129.
\newblock \url{http://stacks.iop.org/1538-3873/129/i=980/a=104502}

\bibitem[{{Kopparapu} {et~al.}(2013){Kopparapu}, {Ramirez}, {Kasting}, {Eymet},
  {Robinson}, {Mahadevan}, {Terrien}, {Domagal-Goldman}, {Meadows}, \&
  {Deshpande}}]{2013ApJ...765..131K}
{Kopparapu}, R.~K., {Ramirez}, R., {Kasting}, J.~F., {et~al.} 2013, \apj, 765,
  131, \dodoi{10.1088/0004-637X/765/2/131}

\bibitem[{Kowalski {et~al.}(2013)Kowalski, L., \& et~al}]{680140843}
Kowalski, L., H.~S., \& et~al, W. J.~P. 2013, ApJS, 207, 15.
\newblock \url{http://stacks.iop.org/0067-0049/207/i=1/a=15}

\bibitem[{Kowalski {et~al.}(2009)Kowalski, Hawley, \& et~al}]{680140841}
Kowalski, A.~F., Hawley, S., \& et~al, E. J.~H. 2009, AJ, 138, 633.
\newblock \url{http://stacks.iop.org/1538-3881/138/i=2/a=633}

\bibitem[{Lacy {et~al.}(1976)Lacy, J, \& S.}]{680140844}
Lacy, C.~H., J, M.~T., \& S., E.~D. 1976, ApJS, 30, 85.
\newblock \url{http://dx.doi.org/10.1086/190358}

\bibitem[{{Lacy} {et~al.}(1976){Lacy}, {Moffett}, \&
  {Evans}}]{1976ApJS...30...85L}
{Lacy}, C.~H., {Moffett}, T.~J., \& {Evans}, D.~S. 1976, \apjs, 30, 85,
  \dodoi{10.1086/190358}

\bibitem[{{Lightkurve Collaboration} {et~al.}(2018){Lightkurve Collaboration},
  {Cardoso}, {Hedges}, {Gully-Santiago}, {Saunders}, {Cody}, {Barclay}, {Hall},
  {Sagear}, {Turtelboom}, {Zhang}, {Tzanidakis}, {Mighell}, {Coughlin}, {Bell},
  {Berta-Thompson}, {Williams}, {Dotson}, \& {Barentsen}}]{lightkurve}
{Lightkurve Collaboration}, {Cardoso}, J.~V.~d.~M., {Hedges}, C., {et~al.}
  2018, {Lightkurve: Kepler and TESS time series analysis in Python},
  Astrophysics Source Code Library.
\newblock \doeprint{1812.013}

\bibitem[{Luger \& Barnes(2015)}]{680140849}
Luger, R., \& Barnes, R. 2015, AsBio, 15, 119.
\newblock \url{http://dx.doi.org/10.1089/ast.2014.1231}

\bibitem[{Maehara {et~al.}(2012)Maehara, T., \& et~al}]{680140850}
Maehara, H., T., S., \& et~al, N.~S. 2012, Natur, 485, 478.
\newblock \url{http://dx.doi.org/10.1038/nature11063}

\bibitem[{Mart{\'{\i}}nez {et~al.}(2020)Mart{\'{\i}}nez, Lopez, Shappee,
  Schmidt, Jayasinghe, Kochanek, Auchettl, \&
  Holoien}]{Rodriguez_Mart_nez_2020}
Mart{\'{\i}}nez, R.~R., Lopez, L.~A., Shappee, B.~J., {et~al.} 2020, The
  Astrophysical Journal, 892, 144, \dodoi{10.3847/1538-4357/ab793a}

\bibitem[{McQuillan {et~al.}(2013)McQuillan, Aigrain, \&
  Mazeh}]{10.1093/mnras/stt536}
McQuillan, A., Aigrain, S., \& Mazeh, T. 2013, Monthly Notices of the Royal
  Astronomical Society, 432, 1203, \dodoi{10.1093/mnras/stt536}

\bibitem[{Mondrik {et~al.}(2018)Mondrik, Newton, Charbonneau, \&
  Irwin}]{Mondrik_2018}
Mondrik, N., Newton, E., Charbonneau, D., \& Irwin, J. 2018, The Astrophysical
  Journal, 870, 10, \dodoi{10.3847/1538-4357/aaee64}

\bibitem[{{Newton} {et~al.}(2016){Newton}, {Irwin}, {Charbonneau},
  {Berta-Thompson}, \& {Dittmann}}]{2016ApJ...821L..19N}
{Newton}, E.~R., {Irwin}, J., {Charbonneau}, D., {Berta-Thompson}, Z.~K., \&
  {Dittmann}, J.~A. 2016, \apjl, 821, L19, \dodoi{10.3847/2041-8205/821/1/L19}

\bibitem[{Notsu {et~al.}(2013)Notsu, Shibayama, Maehara, Notsu, Nagao, Honda,
  Ishii, Nogami, \& Shibata}]{Notsu_2013}
Notsu, Y., Shibayama, T., Maehara, H., {et~al.} 2013, The Astrophysical
  Journal, 771, 127, \dodoi{10.1088/0004-637x/771/2/127}

\bibitem[{{Osten} \& {Wolk}(2015)}]{2015ApJ...809...79O}
{Osten}, R.~A., \& {Wolk}, S.~J. 2015, \apj, 809, 79,
  \dodoi{10.1088/0004-637X/809/1/79}

\bibitem[{Paudel {et~al.}(2018)Paudel, Gizis, Mullan, Schmidt, Burgasser,
  Williams, \& Berger}]{Paudel_2018}
Paudel, R.~R., Gizis, J.~E., Mullan, D.~J., {et~al.} 2018, The Astrophysical
  Journal, 858, 55, \dodoi{10.3847/1538-4357/aab8fe}

\bibitem[{Paudel {et~al.}(2019)Paudel, Gizis, Mullan, Schmidt, Burgasser,
  Williams, Youngblood, \& Stassun}]{Paudel_super}
---. 2019, Monthly Notices of the Royal Astronomical Society, 486, 1438,
  \dodoi{10.1093/mnras/stz886}

\bibitem[{Priest \& Forbes(2002)}]{Priest2002}
Priest, E.~R., \& Forbes, T.~G. 2002, The Astronomy and Astrophysics Review,
  10, 313, \dodoi{10.1007/s001590100013}

\bibitem[{Raetz {et~al.}(2020)Raetz, Stelzer, Damasso, \& Scholz}]{Raetz_2020}
Raetz, S., Stelzer, B., Damasso, M., \& Scholz, A. 2020, Astronomy \&
  Astrophysics, 637, A22, \dodoi{10.1051/0004-6361/201937350}

\bibitem[{{Ricker} {et~al.}(2014){Ricker}, {Winn}, {Vanderspek}, {Latham},
  {Bakos}, {Bean}, {Berta-Thompson}, {Brown}, {Buchhave}, {Butler}, {Butler},
  {Chaplin}, {Charbonneau}, {Christensen-Dalsgaard}, {Clampin}, {Deming},
  {Doty}, {De Lee}, {Dressing}, {Dunham}, {Endl}, {Fressin}, {Ge}, {Henning},
  {Holman}, {Howard}, {Ida}, {Jenkins}, {Jernigan}, {Johnson}, {Kaltenegger},
  {Kawai}, {Kjeldsen}, {Laughlin}, {Levine}, {Lin}, {Lissauer}, {MacQueen},
  {Marcy}, {McCullough}, {Morton}, {Narita}, {Paegert}, {Palle}, {Pepe},
  {Pepper}, {Quirrenbach}, {Rinehart}, {Sasselov}, {Sato}, {Seager},
  {Sozzetti}, {Stassun}, {Sullivan}, {Szentgyorgyi}, {Torres}, {Udry}, \&
  {Villasenor}}]{2014SPIE.9143E..20R}
{Ricker}, G.~R., {Winn}, J.~N., {Vanderspek}, R., {et~al.} 2014, in Society of
  Photo-Optical Instrumentation Engineers (SPIE) Conference Series, Vol. 9143,
  \procspie, 914320, \dodoi{10.1117/12.2063489}

\bibitem[{{Rimmer} {et~al.}(2018){Rimmer}, {Xu}, {Thompson}, {Gillen},
  {Sutherland}, \& {Queloz}}]{2018SciA....4.3302R}
{Rimmer}, P.~B., {Xu}, J., {Thompson}, S.~J., {et~al.} 2018, Science Advances,
  4, eaar3302, \dodoi{10.1126/sciadv.aar3302}

\bibitem[{{Saar} \& {Linsky}(1985)}]{1985BAAS...17..751S}
{Saar}, S.~H., \& {Linsky}, J.~L. 1985, in \baas, Vol.~17, 879

\bibitem[{Schmidt {et~al.}(2014)Schmidt, L., \& et~al}]{680140886}
Schmidt, S.~J., L., P.~J., \& et~al, S. K.~Z. 2014, ApJL, 781, L24.
\newblock \url{http://stacks.iop.org/2041-8205/781/i=2/a=L24}

\bibitem[{Schmidt {et~al.}(2019)Schmidt, Shappee, van Saders, Stanek, Brown,
  Kochanek, Dong, Drout, Frank, Holoien, Johnson, Madore, Prieto, Seibert,
  Seidel, \& Simonian}]{Schmidt_2019}
Schmidt, S.~J., Shappee, B.~J., van Saders, J.~L., {et~al.} 2019, The
  Astrophysical Journal, 876, 115, \dodoi{10.3847/1538-4357/ab148d}

\bibitem[{Segura {et~al.}(2010)Segura, M., andKasting J., \& S.}]{680140862}
Segura, A., M., W.~L., andKasting J., M.~V., \& S., H. 2010, AsBio, 10, 751.
\newblock \url{http://dx.doi.org/10.1089/ast.2009.0376}

\bibitem[{Shappee {et~al.}(2014)Shappee, L., \& et~al}]{680140864}
Shappee, B.~J., L., P.~J., \& et~al, G.~D. 2014, ApJ, 788, 48.
\newblock \url{http://stacks.iop.org/0004-637X/788/i=1/a=48}

\bibitem[{{Shields} {et~al.}(2016){Shields}, {Ballard}, \&
  {Johnson}}]{2016PhR...663....1S}
{Shields}, A.~L., {Ballard}, S., \& {Johnson}, J.~A. 2016, \physrep, 663, 1,
  \dodoi{10.1016/j.physrep.2016.10.003}

\bibitem[{Silverberg {et~al.}(2016)Silverberg, F., \& et~al}]{680140866}
Silverberg, S.~M., F., K.~A., \& et~al, D. J. R.~A. 2016, ApJ, 829, 129.
\newblock \url{http://stacks.iop.org/0004-637X/829/i=2/a=129}

\bibitem[{Simonian {et~al.}(2016)Simonian, Z., \& et~al}]{680140867}
Simonian, G., Z., S.~K., \& et~al, S.~S. 2016, ATel, 8803.
\newblock \url{http://adsabs.harvard.edu/abs/2016ATel.8803....1S}

\bibitem[{Stanek {et~al.}(2013)Stanek, J., \& et~al}]{680140870}
Stanek, K.~Z., J., S.~B., \& et~al, K. C.~S. 2013, ATel, 5276.
\newblock \url{http://adsabs.harvard.edu/abs/2013ATel.5276....1S}

\bibitem[{Tilley {et~al.}(2019)Tilley, Segura, Meadows, Hawley, \&
  Davenport}]{Tilley2019}
Tilley, M.~A., Segura, A., Meadows, V., Hawley, S., \& Davenport, J. 2019,
  Astrobiology, 19, 64, \dodoi{10.1089/ast.2017.1794}

\bibitem[{{Virtanen} {et~al.}(2020){Virtanen}, {Gommers}, {Oliphant},
  {Haberland}, {Reddy}, {Cournapeau}, {Burovski}, {Peterson}, {Weckesser},
  {Bright}, {van der Walt}, {Brett}, {Wilson}, {Jarrod Millman}, {Mayorov},
  {Nelson}, {Jones}, {Kern}, {Larson}, {Carey}, {Polat}, {Feng}, {Moore}, {Vand
  erPlas}, {Laxalde}, {Perktold}, {Cimrman}, {Henriksen}, {Quintero}, {Harris},
  {Archibald}, {Ribeiro}, {Pedregosa}, {van Mulbregt}, \&
  {Contributors}}]{2020SciPy-NMeth}
{Virtanen}, P., {Gommers}, R., {Oliphant}, T.~E., {et~al.} 2020, Nature
  Methods, 17, 261, \dodoi{https://doi.org/10.1038/s41592-019-0686-2}

\end{thebibliography}
\bibliographystyle{aasjournal}



\end{document}